\documentclass[3p,times,preprint]{elsarticle}

\usepackage[export]{adjustbox}

\usepackage{amssymb}
\usepackage{amsmath}
\usepackage{float}
\usepackage{subfig}

\usepackage[figuresright]{rotating}

\begin{document}

\begin{frontmatter}




\title{A conservative implicit-PIC scheme for the hybrid kinetic-ion fluid-electron plasma model on curvilinear meshes}

\author{A.~Stanier, L. Chac\'{o}n}

\address{Applied Mathematics and Plasma Physics, Los Alamos National Laboratory, Los Alamos, NM 87545, United States}

\begin{abstract}
The hybrid kinetic-ion fluid-electron plasma model is widely used to study challenging multi-scale problems in space and laboratory plasma physics. Here, a novel conservative scheme for this model employing implicit particle-in-cell techniques~\cite{stanier19cart} is extended to arbitrary coordinate systems via curvilinear maps from logical to physical space. The scheme features a fully non-linear electromagnetic fomulation with a multi-rate time advance - including sub-cycling and orbit-averaging for the kinetic ions. By careful choice of compatible particle-based kinetic-ion and mesh-based fluid-electron discretizations in curvilinear coordinates, as well as particle-mesh interpolations and implicit midpoint time advance, the scheme is proven to conserve total energy for arbitrary curvilinear meshes. In the electrostatic limit, the method is also proven to conserve total momentum for arbitrary curvilinear meshes. Although momentum is not conserved for arbitrary curvilinear meshes in the electromagnetic case, it is for an important subset of Cartesian tensor-packed meshes. The scheme and its novel conservation properties are demonstrated for several challenging numerical problems using different curvilinear meshes, including a merging flux-rope simulation for a space weather application, and a helical $m=1$ mode simulation for magnetic fusion energy application.\end{abstract}


\end{frontmatter}


\section{Introduction}\label{}


Magnetized plasmas in nature and the laboratory can be inherently multi-scale, with extreme scale disparity between the global system sizes and the ion and electron kinetic orbit scales. From a modeling perspective, the magnetohydrodynamic (MHD) fluid approach is widely used to model the bulk plasma flows and the evolution of the magnetic topology at the largest scales~\cite{priest2014}, whereas the high-fidelity kinetic Vlasov-Maxwell model is used for detailed study of plasma micro-scales~\cite{schindler2006}. However, it is highly desirable to bridge this scale gap for many systems to enable the study of cross-scale coupling. This scale-bridging is likely best accomplished through the combined use of reduced kinetic models (which are of lower fidelity than the full kinetic model yet capture key physics) along with algorithmic advances to step over the stiffest space and time scales in a stable manner.



A promising class of reduced models take a \textit{hybrid} approach, in which some plasma components are treated with a high-fidelity kinetic description, and the remainder using a lower-fidelity reduced fluid approximation~\cite{lipatov02}. Variants include the kinetic-ion fluid-electron model~\cite{byers78,hewett78,winske03}, the fast-particle-kinetic bulk-fluid model~\cite{belova1997,drake2019,holderied2021}, as well as generalized hybrid models that are flexible to different choices of the kinetic and fluid components~\cite{amano2018}. In this paper, we focus on the kinetic-ion fluid-electron model with the kinetic ions being solved with the particle-in-cell (PIC) technique~\cite{birdsall91}, and assume a zero-inertia and scalar pressure electron fluid. This model is widely used to study the macro- to ion-scale coupling in magnetic reconnection~\cite{karimabadi04,stanier15prl,le2019}, collisionless shocks~\cite{winske1985,weidl2016,le2021}, collisionless turbulence~\cite{franci2015,cerri2016,le2018wavelet} as well as global magnetospheric physics~\cite{muller2012,karimabadi14,omidi2017,ng2021bursty}.

Common numerical schemes for this model employ explicit time-stepping. As discussed in Ref.~\cite{winske03}, a key algorithmic difficulty with explicit schemes concerns the use of a static Ohm's law to evaluate the electric field along with a leap-frog advance for the particles. The time advanced electric field is needed to advance the particles, but this in turn depends implicitly on the time advanced particle velocities. To make this explicit, some form of predictor-corrector~\cite{harned82,winske86,kunz13}, moment-based~\cite{winske88,matthews94}, or forward-extrapolation schemes~\cite{fujimoto90,thomas90,karimabadi04} have been used~(see also Refs.~\cite{winske03,karimabadi04} for reviews). These schemes are often highly efficient, but the approximations made are such that the numerical conservation of momentum and energy is violated~\cite{stanier19cart}. A further numerical difficulty with explicit schemes concerns the extremely stiff CFL-condition associated with the quadratic dispersion relation for whistler waves, namely $\Delta t \propto n (\Delta x)^2/B$, where $\Delta t$ is the timestep, $\Delta x$ the grid spacing, $n$ the density and $B$ the magnetic field-strength. This can be extremely stiff in strongly magnetized and near-vacuum regions, and often requires the syb-cycling of the magnetic field-solve~\cite{swift95,matthews94,karimabadi04} or some other special treatment~\cite{hewett80,harned82,amano14} to avoid numerical instability.

Implicit methods have been used to solve the electromagnetic field equations~\cite{hewett80,lipatov02} and, more recently, for the kinetic particle advance. In particular, Ref.~\cite{sturdevant17} explored implicit timestepping for particles featuring sub-stepping and orbit-averaging techniques in an electrostatic $\delta F$ hybrid model. More recently, Ref.~\cite{stanier19cart} proposed a fully implicit scheme for the non-linear electromagnetic kinetic-ion fluid-electron hybrid model, which combines an implicit-midpoint time advance with a cell-centered finite difference spatial discretization, and includes particle sub-cycling and orbit averaging. This latter scheme, which was formulated for uniformly spaced meshes in Cartesian geometry, features simultaneous discrete conservation of momentum and energy. It was demonstrated that this scheme also features favorable stability properties with respect to the finite grid-instability, which can occur in non-conservative schemes when the ratio of the ion-to-electron temperatures is small ($T_i/T_e \ll 1$)~\cite{rambo95,stanier19cart}.

Non-uniform or curvilinear mapped meshes can be used to resolve thin boundary layer features efficiently, or to model boundary-fitted domains that have specific geometries. For the hybrid-PIC model, explicit schemes have been formulated with cylindrical~\cite{hewett80,higginson19}, toroidal~\cite{leblond11phd,sturdevant17}, or spherical meshes~\cite{swift95,dyadechkin13} to study space or laboratory plasmas. Adaptive-Mesh Refinement (AMR) techniques have also been applied to the hybrid-PIC model for global magnetosphere simulations~\cite{muller2011aikef}. The use of non-uniform meshes and/or curvilinear coordinates within PIC methods can be tricky in general and careful choices of compatible grid and particle-based discretizations, as well as particle-grid interpolations, are required to avoid large violations in momentum and energy conservation. One issue concerns the non-physical particle self-forces, which can be amplified in regions of strong mesh gradients to potentially cause artificial particle reflection~\cite{colella10}. A large range of mesh resolutions across the simulation domain can also cause problems both in terms of statistical noise -- due to a different number of macro-particles in each cell -- as well as by the loss in stability or accuracy in coarse grid regions for numerical schemes that are not asymptotic preserving. For the latter, it is well known for standard kinetic Vlasov-Maxwell PIC algorithms~\cite{birdsall91} that significant numerical heating can occur if the electron Debye length $\lambda_{D,e} = v_{\textrm{th},e}/\omega_{pe}$ is not resolved everywhere, where $v_{\textrm{th},e}$ is the electron thermal speed and $\omega_{pe}$ is the plasma frequency. For the quasi-neutral hybrid kinetic-ion fluid-electron PIC model, the effects can be more subtle. It has been recently shown~\cite{stanier20cancel} that significant unphysical wave dispersion errors can occur in hybrid-PIC when combining the use of higher-order particle-mesh interpolation functions with under-resolving the ion skin-depth $d_i=v_A/\Omega_{ci}$, where $v_A$ is the Alfv\'en speed and $\Omega_{ci}$ is the ion gyro-frequency. These errors are caused by the particle-mesh interpolations spreading the electric field that is experienced by the ion particles, and leading to its inexact cancellation when combining the particle-based ion and grid-based electron momentum equations. Remarkably, the source of these errors is removed when using the lowest order \textit{nearest grid-point} (NGP) interpolation functions.

In this paper, we extend the fully implicit scheme of Ref.~\cite{stanier19cart} to treat arbitrary coordinate systems via curvilinear maps from logical to physical space. This novel hybrid-PIC scheme has the following unique conservation properties: 1) it conserves energy numerically for arbitrary curvilinear maps (assuming suitable boundary conditions), 2) in the electrostatic limit it also conserves total linear momentum for general meshes, and 3) for the full electromagnetic model, it features discrete momentum conservation for Cartesian meshes with a tensor-product mesh packing function.  Although the latter property only holds for a narrow subset of meshes, it is notable since the lack of artificial self-forces means that it avoids the issue of spurious particle reflection mentioned above. Section~\ref{sec:model} of the paper reviews the quasi-neutral hybrid kinetic-ion fluid-electron plasma model. We describe our implicit-PIC discretization of the model for curvilinear meshes in Section~\ref{sec:discrete}, and prove its discrete conservation properties in Sec.~\ref{sec:conserve}. In Section~\ref{sec:implement}, we detail our particular numerical implementation and solution strategy for the discrete model equations, and finally, in Section~\ref{sec:verify}, we present numerical example problems that utilize different mesh geometries and verify the novel conservation properties of our scheme.

\section{\label{sec:model}Electromagnetic kinetic-ion fluid-electron hybrid model}

In the electromagnetic and quasi-neutral hybrid kinetic-ion fluid-electron model, ions with species index $s$ are described by the distribution function $f_s(t,\boldsymbol{v},\boldsymbol{x})$ that is found from the Vlasov equation as

\begin{equation}\label{ionvlasov}\partial_tf_s + \boldsymbol{\nabla} \cdot \left(f_s \boldsymbol{v}\right) + \left(q_s/m_s\right)\left(\boldsymbol{E}^* + \boldsymbol{v}\times \boldsymbol{B}\right) \cdot \boldsymbol{\nabla}_v f_s = 0.\end{equation}
Here, $q_s, m_s$ are the ion species charge and mass, respectively, $\boldsymbol{B}$ is the magnetic field, and $\boldsymbol{E}^*$ is the frictionless electric field (see below). The magnetic field is advanced by Faraday's equation as
\begin{equation}\label{continuumfaraday}\partial_t \boldsymbol{B} = - \boldsymbol{\nabla} \times \boldsymbol{E},\end{equation}
where $\boldsymbol{E}$ is the total electric field. The latter is evaluated using an Ohm's law as
\begin{equation}\label{continuumohms}\boldsymbol{E} = \boldsymbol{E}^* + \boldsymbol{F}_{ie} = - \boldsymbol{u}_i\times \boldsymbol{B} + \frac{\boldsymbol{j}\times \boldsymbol{B}}{ne} - \frac{\boldsymbol{\nabla}p_e}{ne} + \boldsymbol{F}_{ie},\end{equation}
where $e$ is the elementary charge, $n = \sum_s (q_s/e) \int f_s d^3v$ is the quasi-neutral density, $\boldsymbol{u}_i = \left(\sum_s q_s \int f_s \boldsymbol{v}d^3v\right)/en$ is the ion current carrying drift velocity, $\boldsymbol{j}=\boldsymbol{\nabla}\times \boldsymbol{B}/\mu_0$ is the current density, $p_e$ is a scalar electron pressure, $\boldsymbol{F}_{ie} = \eta \boldsymbol{j}$ is the resistive friction, and $\eta$ is the plasma resistivity. We note that only the frictionless electric field $\boldsymbol{E}^* = \boldsymbol{E} - \boldsymbol{F}_{ie}$ is used in Eq.~(\ref{ionvlasov}) as is required for momentum conservation, see e.g. Ref.~\cite{stanier19cart}.
 
Finally, the scalar electron pressure is calculated from
\begin{equation}\label{continuumpressure}\frac{1}{\gamma-1}\left[\partial_t p_e + \boldsymbol{\nabla} \cdot  \left(\boldsymbol{u}_e p_e\right)\right] + p_e \boldsymbol{\nabla}\cdot \boldsymbol{u}_e = H_e - \boldsymbol{\nabla}\cdot \boldsymbol{q}_e,\end{equation}
where $\gamma=5/3$ is the ratio of specific heats, $\boldsymbol{u}_e = \boldsymbol{u}_i - \boldsymbol{j}/en$ is the electron bulk velocity, $H_e = \eta j^2$ is the frictional (Joule) heating term, $\boldsymbol{q}_e = - \kappa \boldsymbol{\nabla}(p_e/n)$ is the electron heat flux, and $\kappa$ is the electron heat conductivity.

With suitable choice of boundary conditions, the set of equations (\ref{ionvlasov}-\ref{continuumpressure}) conserves the total mass $\left(\int \sum_s m_s \int f_s d^3v d^3x\right)$, linear momentum $\left(\int \sum_s m_s \int f_s \boldsymbol{v} d^3v d^3x\right)$, and energy

\begin{equation}\int \left[\sum_s \int \tfrac{1}{2} m_s v^2 f_s d^3v + p_e/(\gamma - 1) + B^2/2\mu_0\right] d^3x,\end{equation}
see e.g. Refs.~\cite{lipatov02,stanier19cart} for details.

\section{\label{sec:discrete}Discretized multi-rate scheme on curvilinear mapped meshes}

\subsection{\label{timediscretization}Time discretization}
For the time advance, we employ the second-order implicit-midpoint method. For a variable $\chi$ with known value at timestep $t=n\Delta t$, i.e. $\chi^n$, the new value $\chi^{n+1}$ is found implicitly from $(\chi^{n+1}-\chi^n)/\Delta t = f(\chi^{n+1/2})$ where $\chi^{n+1/2}  = \tfrac{1}{2}(\chi^{n+1} + \chi^n)$. We note that this differs from the Crank-Nicolson method in the discretization of non-linear terms. This timestepping scheme is used here both for the grid-based equations (Sec.~\ref{sec:fielddiscretization}), and for the particle advance (Sec.~\ref{sec:particlediscretization}) where it gives long term accuracy for ion orbit integration (it is symplectic). This method was found to give exact discrete conservation in the Cartesian formulation of this hybrid-PIC kinetic-ion and fluid-electron scheme~\cite{stanier19cart}. 

\subsection{\label{curvilinearoperations}Curvilinear map and basis}

Assuming a static and non-singular curvilinear map from logical space $\xi^{\alpha} = (\xi,\eta,\mu)$ to physical space $\boldsymbol{x} = (x,y,z)$, i.e. $\boldsymbol{x}(\xi^{\alpha})$, the Jacobian of the mapping is defined as $J=|\partial \boldsymbol{x}/\partial \xi^{\alpha}|$, the contravariant basis vectors as $\boldsymbol{\nabla} \xi^{\alpha}$, and the covariant basis vectors as $\partial_\alpha \boldsymbol{x} \equiv \partial \boldsymbol{x}/\partial \xi^{\alpha}$. Then, for a generic vector $\boldsymbol{S}$, the contravariant components are defined as $S^{\alpha} = \boldsymbol{S}\cdot \boldsymbol{\nabla} \xi^{\alpha}$, and the covariant components as $S_{\alpha} = \boldsymbol{S} \cdot \partial_\alpha \boldsymbol{x}$. Indices can be lowered as $S_\alpha = g_{\alpha \beta} S^{\beta}$ via the metric tensor $g_{\alpha \beta} \equiv \partial_{\alpha}\boldsymbol{x}\cdot\partial_\beta\boldsymbol{x}$, and raised as $S^{\alpha} = g^{\alpha \beta} S_{\beta}$ where $g_{\alpha \beta}g^{\beta \gamma} = \delta_{\alpha}^{\gamma}$. \ref{sec:curvilinearidentities} gives some further definitions and vector identities used herein.

\subsection{\label{sec:fielddiscretization}Discretization of the field equations}

In hybrid-PIC discretizations of the model described in Section~\ref{sec:model},  Eqs.~(\ref{continuumfaraday}, \ref{continuumohms}, \ref{continuumpressure}) for $\boldsymbol{B}$, $\boldsymbol{E}$, and $p_e$ are typically solved on a spatial grid. Here, we use a uniform structured grid in logical space with all grid variables $\chi_g \equiv \chi_{ijk}$ defined at the position of the cell centers in logical space $(\xi_i,\eta_j,\mu_k)$ for $i \in [1,N_{\xi}]$, $j\in [1,N_\eta]$, $k \in [1,N_{\mu}]$. The subscript $g$ (shorthand for the 3-dimensional grid index $ijk$) is used to denote grid-based variables herein. The spatial derivatives of these mesh variables are computed using a second-order centered finite-difference approximation, as e.g. $(\partial_\xi \chi)_{g} \equiv (\partial_\xi \chi)_{ijk} = (\chi_{i+1jk}-\chi_{i-1jk})/(2\Delta\xi)$, where $\Delta \xi$ is the uniform grid spacing on the logical mesh. 

The covariant electric field $E_{\alpha,g}^{n+1/2}$ is found via static evaluation of Ohm's law at the $n+1/2$ time level as
\begin{equation}\label{ohms}E_{\alpha,g}^{n+1/2} = -J_{g}\epsilon_{\alpha\beta\gamma}(u^\beta)^{n+1/2}_{g}(B^\gamma)^{n+1/2}_{g} + J_{g}\epsilon_{\alpha\beta\gamma}\frac{(j^\beta)_{g}^{n+1/2}(B^{\gamma})^{n+1/2}_{g}}{en^{n+1/2}_{g}} -\frac{1}{en^{n+1/2}_{g}} \left(\frac{\partial p_e}{\partial \xi^{\alpha}}\right)^{n+1/2}_{g} + \eta (j_\alpha)_g^{n+1/2},\end{equation}
where $\epsilon_{\alpha \beta \gamma}$ is the Levi-Civita symbol. The density and velocity moments of the kinetic ions, $n_{g}$ and $(u^\beta)_{g}$, are defined below. $(B^{\gamma})_{g}$ are the contravariant components of the magnetic field, $(j^{\beta})_{g} = (1/J_{g}) \epsilon^{\beta \sigma \alpha} (\partial_\sigma (g_{\alpha \gamma}B^{\gamma}))_{g}$ are the contravariant components of the current density, and $(p_e)_{g}$ is the electron scalar pressure. 

The contravariant magnetic field components are calculated via Faraday's equation as
\begin{equation}\label{faraday}\frac{(B^{\alpha})^{n+1}_g - (B^{\alpha})^{n}_g}{\Delta t} = - \frac{1}{J_{g}}\epsilon^{\alpha \beta \gamma} (\partial_\beta E_\gamma)^{n+1/2}_{g}.\end{equation}

The scalar electron pressure is updated as
\begin{equation}\label{epress}\frac{1}{(\gamma-1)}\left[\frac{(Jp_e)^{n+1}_{g} - (Jp_e)^n_{g}}{\Delta t} + \partial_\alpha \left(Ju_e^\alpha p_e\right)^{n+1/2}_{g}\right] + (p_e)^{n+1/2}_{g} (\partial_\alpha J u_e^{\alpha})^{n+1/2}_{g} = \eta J_g(j_\alpha j^\alpha)_g^{n+1/2} - \left(\partial_\alpha J_g q_e^\alpha\right)_g^{n+1/2},\end{equation}
where $(u_e^\alpha)_{g} = (u^{\alpha})_{g} - (j^{\alpha})_{g}/en_{g}$ is the electron bulk velocity. Note that we have multiplied Eq.~(\ref{epress}) by $J_g$, which we assume to be constant in time. We note that these choices for discretization of the spatial grid equations are motivated by those used for a non-staggered resistive-MHD  scheme described in Ref.~\cite{chacon04}. 

\subsection{\label{sec:particlediscretization}Discretization of the particle governing equations}
\subsubsection{\label{particlecoordinates}Choice of coordinates: logical in space, Cartesian in velocity-space (hybrid push)}
As discussed above, the real advantage of using curvilinear mapped meshes is to either i) selectively resolve sharp boundary layers, or ii) to use a mesh that aligns with certain features of the simulated problem, such as the magnetic field direction or the shape of the domain boundaries. In contrast, the ion kinetic species are discretized as a collection of Lagrangian phase-space marker particles and are advanced in continuous space. Thus, their accurate integration does not depend on the choice of coordinate system in the same manner. There is freedom to formulate the particle equations of motion in logical space, Cartesian space, or some \textit{hybrid} method of the two. Here, we formulate such a hybrid method in which the particle velocities are represented in Cartesian coordinate directions but the particle positions are updated in logical space. The advantage of the Cartesian velocity update is to avoid the numerical difficulty of additional ficticious force terms in the particle equations of motion, while the logical position update allows the efficient tracking of the particle positions within curved cells (logical cells are regular cubes). 

\subsubsection{\label{subcycle}Multi-rate time integration: sub-cycling of ion orbits}

 As in Ref.~\cite{stanier19cart}, multi-rate time integration is used where the particles can have smaller timesteps than are used to solve Faraday's equation~(\ref{faraday}) and the electron pressure equation~(\ref{epress}). The benefit of this sub-cycling is to give accurate integration of particle orbits in regions of strong electromagnetic fields, or within small cells, while avoiding the need to solve the full non-linear system at each sub-step. This is particularly advantageous for many problems of interest where the characteristic frequency $\omega \ll \Omega_{ci}$, where the particle sub-step can be chosen to follow the fast gyro-frequency $\Omega_{ci}$ and the macro-scale timestep can be chosen to follow $\omega$. Each particle with index $p$ at substep $\nu$ is advanced in time by $\Delta \tau_p^{\nu}$, which is calculated adaptively using local error estimation~\cite{chen15,stanier19cart}. The number of substeps $N_{\nu,p}$ for each particle is such that $\sum_{\nu=0}^{N_{\nu,p} -1}\Delta \tau_p^{\nu}  = \Delta t$.

\subsubsection{Particle push: equations of motion}

The logical-space position of each particle, $\xi^{\alpha}_p$, is updated in time by the substep $\Delta \tau_p^{\nu}$ as
\begin{equation}\label{positioneom}\frac{\left(\xi^\alpha_p\right)^{\nu+1} - \left(\xi^\alpha_p\right)^{\nu}}{\Delta \tau_p^{\nu}} = \boldsymbol{v}_p^{\nu+1/2}\cdot \left(\boldsymbol{\nabla}\xi^\alpha_p\right)^{\nu+1/2},\end{equation}
where $\boldsymbol{v}_p$ is the Cartesian particle velocity vector and the covariant basis vector $\left(\boldsymbol{\nabla}\xi^\alpha_p\right)$ is evaluated at the particle position. To calculate this basis function numerically, we first construct a locally continuous map $\boldsymbol{x}(\xi^{\alpha}_p)=\sum_g\boldsymbol{x}_g S_2(\xi^\alpha_{g} - \xi^{\alpha}_p) \equiv \sum_{ijk}\boldsymbol{x}_{ijk} S_2(\xi_i - \xi_p)S_2(\eta_j - \eta_p)S_2(\mu_k - \mu_p)$, using the tensor product of second-order B-spline shape functions. Then the following identity is used, written for the $\xi$-component ($\alpha=1$) as
\begin{equation}\boldsymbol{\nabla} \xi_p = \frac{1}{J_p} \frac{\partial \boldsymbol{x}}{\partial \eta}\Bigg|_p \times  \frac{\partial \boldsymbol{x}}{\partial \mu}\Bigg|_p,\end{equation}
where the partial derivatives are calculated using the analytical derivatives of the shape functions, which themselves are combinations of lower-order B-splines~\cite{birdsall91}.

The Cartesian particle velocity vector $\boldsymbol{v}_p$ is updated as
\begin{equation}\label{velocityeom}\frac{\boldsymbol{v}_p^{\nu+1}-\boldsymbol{v}_p^{\nu}}{\Delta \tau_p^{\nu}} = \frac{q_p}{m_p}\left(\boldsymbol{E}^{*,\nu+1/2}_p + \boldsymbol{v}_p^{\nu+1/2}\times \boldsymbol{B}_p^{\nu+1/2}\right),\end{equation}
using the Cartesian electromagnetic field vectors at the particle position.

\subsection{Particle-mesh interpolations}

To close the set of equations, electromagnetic fields are scattered from the spatial grid to the particle positions, and moments of the particle distributions are gathered for use in Eqs.~(\ref{ohms}-\ref{epress}). For the momentum conservation properties of our scheme described in Sections~\ref{electrostaticmomentum}-\ref{planargrids}, we find it necessary to scatter the Cartesian components of the electromagnetic fields to the particle positions as
\begin{equation}\label{scatterE}\boldsymbol{E}_p^{*,\nu+1/2} = \sum_{g} \boldsymbol{E}^{*,n+1/2}_{g} S(\xi^\alpha_{g} - (\xi^{\alpha})_p^{\nu+1/2}),\end{equation}
\begin{equation}\label{scatterB}\boldsymbol{B}_p^{\nu+1/2} = \sum_{g} \boldsymbol{B}^{n+1/2}_{g} S(\xi^\alpha_{g} - (\xi^{\alpha})_p^{\nu+1/2}),\end{equation}
where $S(\xi^\alpha_{g} - \xi^{\alpha}_p) \equiv S_m(\xi_i - \xi_p)S_m(\eta_j - \eta_p)S_m(\mu_k - \mu_p)$ is the tensor product of $m$-th order B-spline shape functions. The electromagnetic fields are assumed to be constant in time over the particle substeps (so are given superscript $n+1/2$), such that the change in force occurs only due to the change in particle substep position. The Cartesian vector values of the fields are found on the spatial mesh as $\boldsymbol{E}_{g} = E_{\alpha,g} \left(\boldsymbol{\nabla} \xi^\alpha \right)_{g}$, $ \boldsymbol{B}_{g} = (B^\alpha)_{g} \left(\frac{\partial \boldsymbol{x}}{\partial \xi^\alpha}\right)_{g}$ prior to interpolation. 


The moments are gathered using the orbit-averaging technique~\cite{cohen82}, where a contribution to the moments at each sub-step is weighted by the fraction of the macro-scale timestep, $\Delta \tau_p^{\nu}/\Delta t$. This orbit averaging technique increases the number of statistical samples in the calculation of the moments, thus reducing the noise. Also, as shown below, it is required to give discrete conservation properties in the presence of subcycling. The quasi-neutral density is gathered as
\begin{equation}\label{densitymoment} n_{g}^{n+1/2} = \frac{1}{J_{g}\Delta V}\sum_p \frac{q_p}{e} w_p \frac{1}{\Delta t} \sum_{\nu=0}^{N_{\nu p} -1}S(\xi^\alpha_{g} - (\xi^{\alpha})_p^{\nu+1/2}) \Delta \tau_p^\nu,\end{equation}
where $w_p$ is the macro-particle weight, which we assume to be constant in time for each particle, and $\Delta V = \Delta \xi \Delta \eta \Delta \mu$ is the constant logical cell volume. The Cartesian bulk momentum is gathered as
\begin{equation}\label{currentmoment} (n\boldsymbol{u})_{g}^{n+1/2} = \frac{1}{J_{g}\Delta V}\sum_p \frac{q_p}{e} w_p \frac{1}{\Delta t} \sum_{\nu=0}^{N_{\nu p} -1} S(\xi^\alpha_{g} - (\xi^{\alpha})_p^{\nu+1/2}) \boldsymbol{v}_p^{\nu+1/2} \Delta \tau_p^\nu.\end{equation}
The contravariant component used in the field and moment equations is then found as $(nu^{\alpha})_{g}^{n+1/2} = (n\boldsymbol{u})_{g} \cdot (\boldsymbol{\nabla}\xi^{\alpha})_{g}$.

\section{\label{sec:conserve}Conservation properties of the discrete model}
\subsection{\label{energyconsv}Energy conservation}
\subsubsection{Kinetic energy of the ion particles}
We define the discrete kinetic energy of the ion particles at time level $(n+1)\Delta t$ as $K_i^{n+1} = \sum_p w_p m_p (\boldsymbol{v}_p^{n+1})^2/2$. Noting that $(n+1)\Delta t = n\Delta t + \sum_{\nu=0}^{N_{\nu,p}-1} \Delta \tau_p^{\nu}$, then the change in kinetic energy over the macro timestep is


\begin{equation}\frac{K_i^{n+1} - K_i^{n}}{\Delta t} = \sum_p w_p m_p \sum_{\nu=0}^{N_\nu -1} \frac{(\boldsymbol{v}_p^{\nu+1})^2 - (\boldsymbol{v}_p^{\nu})^2}{2\Delta t}=\sum_p w_p m_p \sum_{\nu=0}^{N_\nu -1} \boldsymbol{v}_p^{\nu+1/2} \cdot \left(\frac{\boldsymbol{v}_p^{\nu+1} - \boldsymbol{v}_p^{\nu}}{\Delta \tau_p^{\nu}}\right) \frac{\Delta \tau_p^{\nu}}{\Delta t}.\end{equation}
Then, using Eqs.~(\ref{velocityeom},\ref{scatterE}, and \ref{currentmoment}),
\begin{equation}\frac{K_i^{n+1} - K_i^{n}}{\Delta t}=\sum_p w_p q_p \sum_{\nu=0}^{N_\nu -1}\boldsymbol{v}_p^{\nu+1/2}\cdot \boldsymbol{E}_p^{*,\nu+1/2} \frac{\Delta \tau_p^{\nu}}{\Delta t}=\sum_{g}  e(Jn\boldsymbol{u})_{g}^{n+1/2} \cdot \boldsymbol{E}^{*, n+1/2}_{g} \Delta V. \end{equation}
Finally, re-writing this in terms of curvilinear components gives
\begin{equation}\frac{K_i^{n+1} - K_i^{n}}{\Delta t}=\sum_{g}  e(nu^{\alpha})_{g}^{n+1/2}(E_{\alpha}^*)^{n+1/2}_{g} J_{g}\Delta V. \nonumber\end{equation}

\subsubsection{Magnetic energy on the discrete mesh}

The total magnetic energy is defined as $W_B^{n+1} = \sum_{g} (B^\alpha)^{n+1}_{g} (B_{\alpha})^{n+1}_{g} J_{g} \Delta V/(2\mu_0)$. The rate of change over a macro-scale timestep can then be written as
\begin{equation}\frac{W_B^{n+1} - W_B^n}{\Delta t} = \sum_g \frac{(B^\alpha)^{n+1}_{g}(B_\alpha)^{n+1}_g - (B^\alpha)_{g}^{n}(B_\alpha)_g^{n}}{2\mu_0 \Delta t}J_g\Delta V = \sum_{g} (g_{\alpha \beta})_{g}\left[\frac{(B^\alpha)_{g}^{n+1}-(B^\alpha)_{g}^{n}}{\mu_0\Delta t}\right] (B^{\beta})_{g}^{n+1/2} J_{g} \Delta V.\end{equation}
Using Eq.~(\ref{faraday}),
\begin{equation}\frac{W_B^{n+1} - W_B^n}{\Delta t} = \sum_{g} (g_{\alpha \beta})_{g} \left[-\epsilon^{\alpha \gamma \delta} (\partial_\gamma E_\delta)_{g}^{n+1/2}\right](B^{\beta})_{g}^{n+1/2} \Delta V. \end{equation}

For suitable boundary conditions, such as periodic, it can be shown that $\sum_g \chi_g (\partial_\alpha \phi_g) = - \sum_g (\partial_\alpha \chi_g) \phi_g$, which is a discrete integration by parts. Using this relation gives
\begin{equation}\frac{W_B^{n+1} - W_B^n}{\Delta t} = \sum_{g}\epsilon^{\alpha \gamma \delta} (\partial_\gamma g_{\alpha \beta} B^\beta)_{g}^{n+1/2} (E_\delta)_{g}^{n+1/2} \Delta V. \end{equation}
Finally, using the definition of the contravariant current density from Sec.~\ref{sec:fielddiscretization} gives
\begin{equation}\frac{W_B^{n+1} - W_B^n}{\Delta t} =\sum_{g} - (j^\delta)_{g}^{n+1/2} (E_{\delta})_{g}^{n+1/2} J_{g} \Delta V.  \end{equation}

\subsubsection{Conservation of total energy}

The total electron kinetic energy (in the absence of electron inertia) is given by $K_e^n = \sum_{g} (p_{e})_g^{n+1} J_{g} \Delta V/(\gamma - 1)$. Assuming suitable boundary conditions so that the flux terms can be neglected, the total energy change
\begin{align}\frac{(K_i + W_B + K_e)^{n+1} - (K_i + W_B + K_e)^n}{\Delta t} = \sum_{g} &\left[e(nu^{\alpha})_{g}^{n+1/2}(E_{\alpha}^*)^{n+1/2}_{g}  - (j^\alpha)_{g}^{n+1/2} (E_{\alpha})_{g}^{n+1/2} + \eta (j_\alpha j^\alpha)_g^{n+1/2}\right]J_g\Delta V\\
-&\left[(p_e)^{n+1/2}_{g} (\partial_\alpha J u_e^{\alpha})^{n+1/2}_{g}\right]\Delta V .\end{align}
Then, using Eq.~(\ref{ohms}) for the first and second terms on the right hand side and using the discrete integration by parts for the final term, this gives
\begin{align}\frac{(K_i + W_B + K_e)^{n+1} - (K_i + W_B + K_e)^n}{\Delta t} &= \sum_{g}  \left[-(u^{\alpha})_{g}^{n+1/2}+\frac{(j^\alpha)_{g}^{n+1/2}}{en_g^{n+1/2}} + (u_e^{\alpha})_g^{n+1/2}\right]\left(\frac{\partial p_e}{\partial \xi^\alpha}\right)_g^{n+1/2}J_g \Delta V\\
&=0,\end{align}
using the definition of the electron bulk velocity.

\subsection{Linear momentum}

The discrete linear momentum is defined as $\boldsymbol{P}^{n+1} = \sum_p w_p m_p \boldsymbol{v}_p^{n+1}$. The rate of change over a macro-scale timestep can then be written as

\begin{equation}\frac{\boldsymbol{P}^{n+1} - \boldsymbol{P}^n}{\Delta t} = \sum_p w_pm_p \sum_{\nu=0}^{N_\nu -1} \left(\frac{\boldsymbol{v}_p^{\nu+1} - \boldsymbol{v}_p^{\nu}}{\Delta \tau^\nu_p}\right) \frac{\Delta \tau^\nu_p}{\Delta t}. \end{equation}
Using Eqs.~(\ref{velocityeom}, ~\ref{scatterE},~\ref{scatterB},~\ref{densitymoment} and~\ref{currentmoment}), this can be written as
\begin{align}\frac{\boldsymbol{P}^{n+1} - \boldsymbol{P}^n}{\Delta t} &= \sum_p w_p q_p \frac{1}{\Delta t} \sum_{\nu=0}^{N_\nu -1} \left(\boldsymbol{E}_p^{\nu+1/2} + \boldsymbol{v}_p^{\nu+1/2} \times \boldsymbol{B}_p^{\nu+1/2}\right) \Delta \tau^\nu_p \\ \nonumber
&= \sum_{g} e\left[n_{g}^{n+1/2} \boldsymbol{E}_{g}^{n+1/2} + (n \boldsymbol{u})_{g}^{n+1/2} \times \boldsymbol{B}_{g}^{n+1/2}\right]J_g\Delta V,\end{align}
which can be written in curvilinear form as 
\begin{equation}\frac{\boldsymbol{P}^{n+1} - \boldsymbol{P}^n}{\Delta t}= \sum_{g} e\left[n_{g}^{n+1/2} (E_\alpha)_{g}^{n+1/2} + J_g \epsilon_{\alpha \beta \gamma} (nu^\beta)_{g}^{n+1/2}(B^{\gamma})_{g}^{n+1/2}  \right]\left(\boldsymbol{\nabla} \xi^\alpha\right)_g J_g\Delta V.\end{equation}
Then, using the covariant form of Ohms law from Eq.~(\ref{ohms}), and cancelling the convective electric field terms gives
\begin{equation}\frac{\boldsymbol{P}^{n+1} - \boldsymbol{P}^n}{\Delta t}= \sum_{g} \left[J_g \epsilon_{\alpha \beta \gamma} \left(j^\beta\right)^{n+1/2}_g \left(B^\gamma\right)^{n+1/2}_g - \frac{\partial p_e}{\partial \xi^\alpha}  \right]\left(\boldsymbol{\nabla} \xi^\alpha\right)_g J_g\Delta V.\end{equation}
Finally, using the definition of the current density $\left(j^\beta\right)_g^{n+1/2}$ and the contraction of indices for the product of Levi-Cevita symbols $\epsilon_{\alpha \beta \gamma}\epsilon^{\beta \sigma \kappa} = -(\delta_\alpha^\sigma \delta_\gamma^\kappa - \delta_\alpha^\kappa \delta_\gamma^\sigma)$ gives
\begin{equation}\label{linearmomentumeqn}\frac{\boldsymbol{P}^{n+1} - \boldsymbol{P}^n}{\Delta t}= \sum_{g} \left[\left(\partial_\gamma B_\alpha\right)_g^{n+1/2} \left(B^\gamma \right)_g^{n+1/2} - \left(\partial_\alpha B_\gamma\right)_g^{n+1/2} \left(B^\gamma \right)_g^{n+1/2} - (\partial_\alpha p_e)_g^{n+1/2}\right]\left(\boldsymbol{\nabla} \xi^\alpha\right)_g J_g\Delta V.\end{equation}


\subsubsection{Linear momentum conservation for general curvilinear meshes in the electrostatic limit}\label{electrostaticmomentum}

Firstly, we consider the electrostatic limit for which the first two terms within the square brackets of Eq.~(\ref{linearmomentumeqn}) are zero. The total linear momentum is then given by the term:

\begin{equation}\frac{\boldsymbol{P}^{n+1} - \boldsymbol{P}^n}{\Delta t} = \sum_{g} -J_{g}(\boldsymbol{\nabla} \xi^{\alpha})_{g}  \left(\frac{\partial p_e}{\partial \xi^{\alpha}}\right)_{g}^{n+1/2} \Delta V.\end{equation} 
Using the discrete integration by parts with suitable (e.g., periodic) boundary conditions, this can be written as
\begin{equation}\label{dJcov}\frac{\boldsymbol{P}^{n+1} - \boldsymbol{P}^n}{\Delta t}=\sum_{g} (p_e)_g^{n+1/2} \left(\partial_\alpha J_{g}(\boldsymbol{\nabla} \xi^{\alpha})_{g}\right)_g  \Delta V.\end{equation} 

As discussed in detail by Ref.~\cite{liseikin2017}, the continuum version of mesh quantity term inside of the large bracket is identically equal to zero for any type of curvilinear map. We have calculated this expression numerically, using our definitions of the covariant derivative and basis functions above, and confirmed that the discrete form of this expression 
\begin{equation}\label{basicidentity}\left(\partial_\alpha J_g \left(\boldsymbol{\nabla} \xi^{\alpha}\right)_g\right)_g = 0\end{equation}
holds locally at every point in our simulation domain down to numerical round-off (at double precision). We have confirmed this result holds even for non-orthogonal meshes.  Thus, in the electrostatic limit, we have that
\begin{equation}\frac{\boldsymbol{P}^{n+1} - \boldsymbol{P}^n}{\Delta t} = 0.\end{equation}
Section~\ref{ESiaw} demonstrates a numerical example of an electrostatic problem that confirms the conservation of total momentum for a non-orthogonal sinusoidal mesh. 


\subsubsection{\label{planargrids}Linear momentum conservation for tensor-packed (Cartesian) meshes in the full electromagnetic model}

Next we consider the electromagnetic terms, which are the first two terms in the square brackets of Eq.~(\ref{linearmomentumeqn}). We find that both of these terms may be non-zero for general curvilinear meshes such as, for example, non-orthogonal meshes or orthogonal meshes in cylindrical or spherical coordinate systems. However, we will proceed to show that they are equal to zero for the subset of Cartesian tensor-packed meshes that are defined by the invertable map $\boldsymbol{x}(\xi^{\alpha}) = (x(\xi), y(\eta), z(\mu))$. This is a incremental generalization of the result from Ref.~\cite{stanier19cart}, where momentum conservation was shown for this electromagnetic model on a \textit{uniform} Cartesian mesh. However, it is an important generalization as this form of mesh is frequently used to study multi-scale problems for which localized regions of the domain require finer resolution, such as the example discussed in Section~\ref{islandcoal}. The conservation of momentum for this type of mesh eliminates unphysical particle self-forces that can otherwise lead to artificial particle reflection in regions of strong mesh gradients, see Ref.~\cite{colella10}.

We firstly note that in the case of $\gamma = \alpha$ the two terms trivially cancel and thus we only need consider the components with $\gamma \neq \alpha$. Writing the summation notation explicitly and dropping the temporal indices (which are all equal) for simplicity of notation, the remaining components can be written as
\begin{equation} \label{sumEMmomentum}\sum_g \sum_\alpha J_g \left(\boldsymbol{\nabla} \xi^\alpha\right)_g \Delta V\sum_{\gamma \neq \alpha} \left[\left(\partial_\gamma B_\alpha\right)_g \left(B^\gamma \right)_g -  \left(\partial_\alpha B_\gamma\right)_g \left(B^\gamma\right)_g\right].\end{equation}


We consider these two terms separately, starting with the second term. We multiply the local value of the expression in the square brackets by the identity transformation $\left(\frac{\partial \boldsymbol{x}}{\partial \xi^\gamma}\right)_g\cdot  \left(\boldsymbol{\nabla} \xi^\gamma\right) \equiv 1$, as
\begin{equation}\sum_{\gamma \neq \alpha}\left(B^\gamma\right)_g \left(\partial_{\alpha} B_\gamma\right)_g \left(\frac{\partial \boldsymbol{x}}{\partial \xi^\gamma}\right)_g\cdot  \boldsymbol{\nabla} \left(\xi^\gamma\right)_g.\end{equation}
Then, using the map for the Cartesian tensor-packed mesh stated above, it can be straightforwardly shown that the quantity $(\boldsymbol{\nabla} \xi^\gamma)_g$ is independent of $\alpha$ for $\gamma \neq \alpha$ (this step is not true for a general curvilinear mesh), and so it can be taken inside the convective derivative term to give
\begin{equation}\sum_{\gamma \neq \alpha} \left(B^\gamma\right)_g \left(\frac{\partial \boldsymbol{x}}{\partial \xi^\gamma}\right)_g \cdot \left(\partial_{\alpha} B_\gamma \boldsymbol{\nabla} \xi^\gamma\right)_g = \sum_{\gamma \neq \alpha} \boldsymbol{B}_g \cdot \left(\partial_{\alpha} \boldsymbol{B}\right)_g = \sum_{\gamma \neq \alpha} \left(\partial_\alpha \frac{B^2}{2}\right)_g,\end{equation}
where $\boldsymbol{B}$ is the Cartesian magnetic field vector. In the final step, we have re-written the expression as the derivative of a particular flux that satisfies the discrete chain rule \textit{locally} and is conservative. It can be shown that, for example, the definition of the ZIP flux satisfies this property, see Refs.~\cite{hirt68,chacon04}. Using this expression, the full second term in Eq.~(\ref{sumEMmomentum}) can be written as
\begin{equation} \sum_g \sum_\alpha J_g \left(\boldsymbol{\nabla} \xi^\alpha\right)_g \Delta V\sum_{\gamma \neq \alpha} \left(\partial_\alpha \frac{B^2}{2}\right)_g.\end{equation}
Then, if we consider the case of $\alpha=1$ without loss of generality, the term  $J_g\left(\boldsymbol{\nabla} \xi\right)_g$ can be written using the definition of the basis functions as
\begin{equation} J_g\left(\boldsymbol{\nabla} \xi\right)_g = \frac{ \partial \boldsymbol{x}}{\partial \eta}\Bigg|_g \times \frac{ \partial \boldsymbol{x}}{\partial \mu}\Bigg|_g. \end{equation} 
For a Cartesian tensor-packed mesh, this is independent of the coordinate $\xi$ and thus the term can be taken inside the convective derivative term to give
\begin{equation} \sum_g \sum_\alpha \Delta V\sum_{\gamma \neq \alpha} \left(\partial_\alpha J_g \left(\boldsymbol{\nabla} \xi^\alpha\right)_g  \frac{B^2}{2}\right)_g = 0.\end{equation}

For the first term in Eq.~(\ref{sumEMmomentum}), we assume a suitably defined (ZIP) flux and rewrite this using the discrete local chain rule as
\begin{equation}\sum_g \sum_\alpha \left(\boldsymbol{\nabla} \xi^\alpha\right)_g \Delta V\sum_{\gamma \neq \alpha} \left[\left(\partial_\gamma J_g B_\alpha B^\gamma\right)_g - B_{\alpha} \underbrace{\left(\partial_\gamma J_gB^\gamma\right)}_{=0}\right],\end{equation}
where the second term in this expression is zero due to the solenoidal property of the magnetic field, which holds locally for our discretization for general curvilinear meshes - see Ref.~\cite{chacon04}. Then, as before, noting that $(\boldsymbol{\nabla} \xi^\alpha)_g$ is independent of $\gamma$ for $\gamma \neq \alpha$, this can be written as a pure derivative of fluxes which is conservative, i.e.

\begin{equation}\sum_g \sum_\alpha \Delta V\sum_{\gamma \neq \alpha} \left[\left(\partial_\gamma J_g \left(\boldsymbol{\nabla} \xi^\alpha\right)_g B_\alpha B^\gamma\right)_g\right] = 0.\end{equation}

The numerical confirmation of momentum conservation in Cartesian tensor-packed meshes is demonstrated in the numerical examples of Sections~\ref{whistler} and ~\ref{islandcoal}.

\section{\label{sec:implement}Numerical implementation}

\subsection{Iterative solution}

As in Ref.~\cite{stanier19cart}, our implementation solves for the magnetic vector potential $(A_\alpha)^{n+1}_g$ rather than the magnetic field formulation $(B^{\alpha})_g^{n+1}$ used in Eq.~(\ref{faraday}). The Weyl gauge is chosen with electrostatic potential $\phi=0$, such that $(A_\alpha)^{n+1}_g$ is found from the solution of
\begin{equation}\label{aform}\frac{(A_\alpha)_g^{n+1} - (A_\alpha)_g^{n}}{\Delta t} = - \left(E_\alpha\right)_g^{n+1/2}.\end{equation}
We note that all of the conservation properties of Sec.~\ref{sec:conserve} also hold for the $\boldsymbol{A}$-formulation with the definition of $(B^{\alpha})_g = \epsilon^{\alpha \beta \gamma} \left(\partial_\beta A_\gamma\right)_g/J_g$. 

The non-linear algebraic equations~(\ref{aform},~\ref{epress}) can be written in the form $\boldsymbol{G}(\boldsymbol{y}^{n+1}) = \boldsymbol{0}$ for the solution vector $\boldsymbol{y}^{n+1} = \left((A_\alpha)_g^{n+1}, (p_e)_g^{n+1}\right)$.  The full details of the method of solution closely follow Ref.~\cite{stanier19cart}. The system is approximately inverted by using a Jacobian-free Newton Krylov (JFNK) method~\cite{knoll04} using flexible-GMRES~\cite{saad93} as a linear solver. The outer Newton iteration is iterated until the residual is converged to a user specified non-linear tolerance $\epsilon_t$. Note that the particle quantities $(\xi^\alpha_p, \boldsymbol{v}_p)$ do not enter the residual directly; rather, the particles are advanced by Eqs.~(\ref{positioneom},~\ref{velocityeom}) and moments gathered by Eqs.~(\ref{densitymoment},~\ref{currentmoment}) in each residual evaluation. To do this, a Picard-iterated implicit Boris method is used to solve Eqs.~(\ref{positioneom},~\ref{velocityeom}) - see Ref.~\cite{stanier19cart}. 

Effective preconditioning plays a primary role in the efficiency of methods such as these. However, the discussion of a suitable physics-based preconditioner is outside of the scope of the present paper, and will be documented elsewhere. For these results, we do not perform preconditioning and, as such, we typically use a small timestep and do not consider performance data for the results presented here.

\subsection{Normalization}\label{sec:normalize}

The discrete model is implemented in normalized form using natural magnetized ion units. For a typical magnetic field strength $B_0$, density $n_0$, and ion mass $m_0$, velocities are normalized by the Alfv\'en speed $v_0 = v_{A0} = B_0/\sqrt{m_0 \mu_0 n_0}$, times by the inverse cyclotron period $t_0 = \Omega_{c0}^{-1} = m_0/eB_0$, and lengths by the ion skin depth $L_0 = d_{i0} = v_{A0}/\Omega_{c0}$. The different species of kinetic ions are distinguished by their normalized charge $Z_s = q_s/e$,  mass $M_s = m_s/m_0$, and density $N_s = n_{s0}/n_0$. If they are initialized with a Maxwellian distribution function with temperature $T_{s0}$, we define the thermal velocity as $v_{Ts0} = \sqrt{T_{s0}/m_s}$ and plasma-beta $\beta_{s0} = 2M_s N_s (v_{Ts}/v_{A0})^2 = 2\mu_0 n_{s0}T_{s0}/B_0^2$. The electron sound speed is defined as $C_{s} = \sqrt{T_{e0}/m_0}$ and the ratio of ion to electron temperatures is $\tau_s = T_{s0}/T_{e0}$.

\section{\label{sec:verify}Numerical Verification}\label{numerical}

\subsection{Electrostatic Ion Acoustic Wave: Momentum conservation}\label{ESiaw}

\begin{figure}
\begin{center}
\includegraphics[trim={0cm 0.2cm 1cm 3.2cm},clip,width=0.6\textwidth]{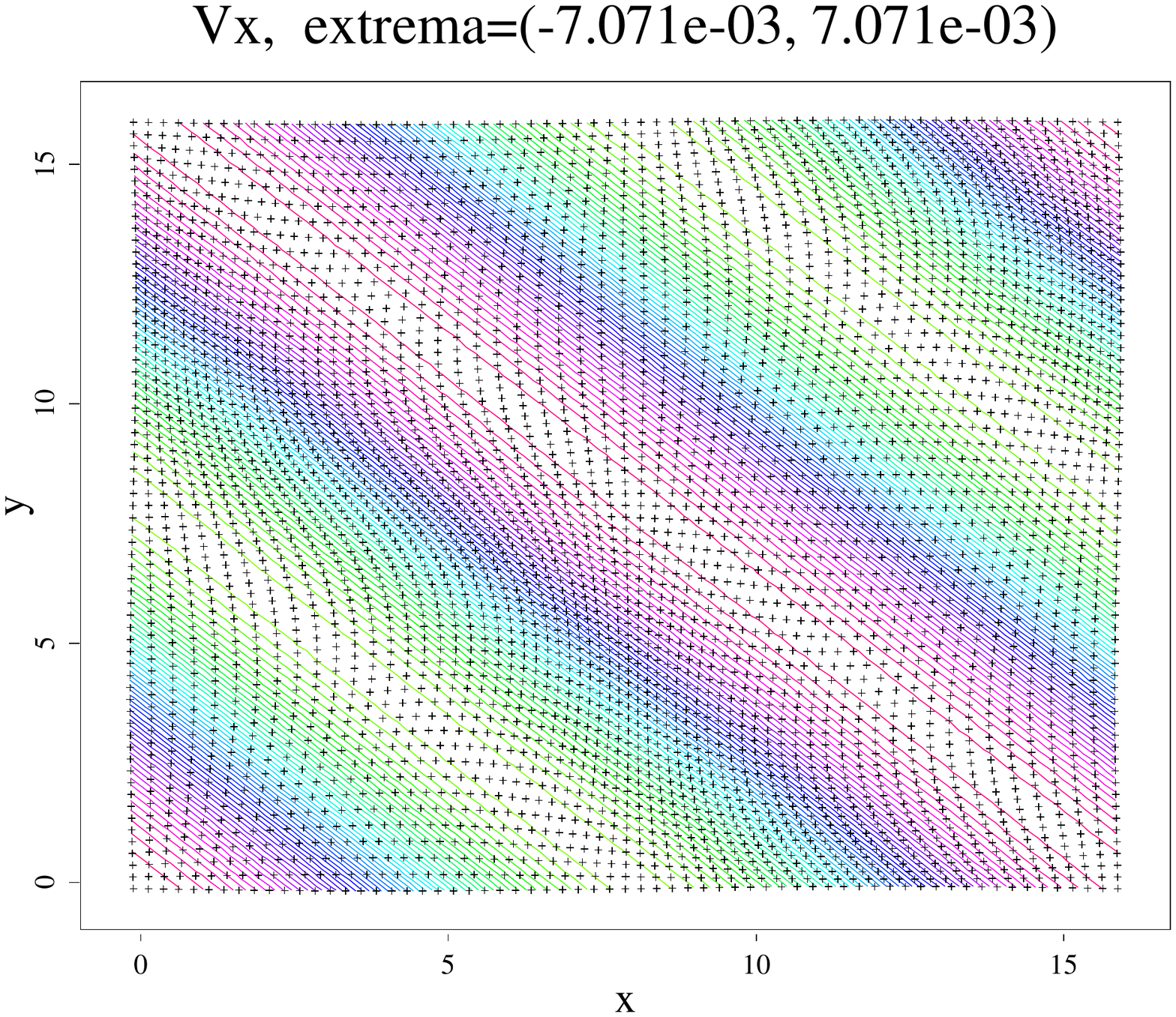}
\caption{\label{iawgrid}Non-orthogonal sinusoidal mesh used for electrostatic ion acoustic wave problem. The colored lines show contours of the Cartesian ion velocity moment $u_{ix}$, and the black $+$ symbols show the centers of the spatial cells. The mesh map is given by Eqs.~(\ref{sinmesheq1}-\ref{sinmesheq2}) with distortion factor $\sigma=1$.}
\end{center}
\end{figure}

In the first numerical example, we consider the ion Landau damping of an ion acoustic wave using a 2D non-orthogonal curvilinear mesh. This is an electrostatic problem and the motivation is to numerically verify the novel conservation properties of the scheme for this limit (momentum and energy conservation, see Sec.~\ref{electrostaticmomentum}) for non-orthogonal mapped meshes. The physical problem set-up considered is similar to that described in Ref.~\cite{stanier19cart}. The simulation is initialized with a single ion species with $Z_i = M_i = 1$ with a Maxwellian velocity function with thermal speed $v_{Ti0}=\sqrt{1/3}$. We set $\gamma = 5/3$ and use a temperature ratio of $\tau = 0.2$. The ions are given a perturbation to their velocity such that the initial bulk velocity is given by $\boldsymbol{\delta u}_i = 0.01\cos{\left(k_x x + k_y y\right)} \left(\boldsymbol{\hat{x}} + \boldsymbol{\hat{y}}\right)$ with $k_x=k_y=\pi/8$. We use a non-orthogonal sinusoidal mesh with periodic boundary conditions for this problem, specified by the map
\begin{equation}\label{sinmesheq1}x(\xi,\eta) = \xi +  \sigma\sin{(2\pi \xi/L_x)}\sin{(2\pi \eta/L_y)}, \end{equation}
\begin{equation}\label{sinmesheq2}y(\xi, \eta) = \eta +  \sigma \sin{(2\pi \xi/L_x)}\sin{(2\pi \eta/L_y)}.\end{equation}
Here, the parameter $\sigma$ describes the distortion of the mesh. Fig.~\ref{iawgrid} shows the initial conditions for the $\boldsymbol{\hat{x}}$-component of the ion velocity moment, along with the spatial mesh used for $\sigma = 1$.  

Additional numerical parameters used are $64\times 64$ cells, with $5\times 10^4$ particles/cell initialized with a quasi-quiet start~\cite{sydora1999} from a low discrepancy Hammersley sequence~\cite{hammersley2013} to reduce noise. We use second-order quadratic-spline shape functions and apply two passes of conservative binomial smoothing, see~\ref{conservativesmoothing}. The timestep used is $\Delta t =0.02$, and the non-linear tolerance used is $\epsilon_t = 10^{-12}$.

\begin{figure}
\begin{center}
\subfloat[]{\includegraphics[width=0.45\textwidth]{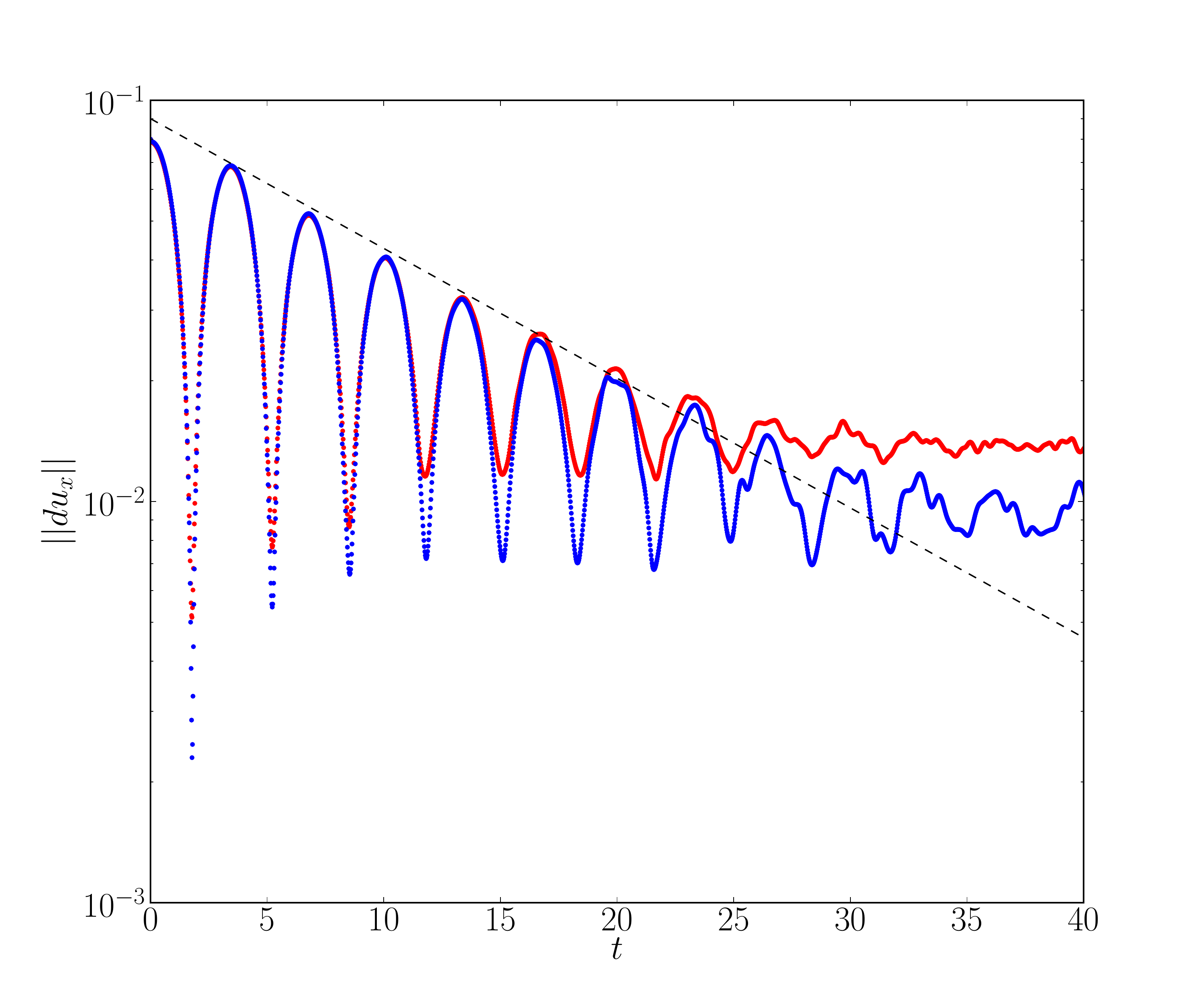}}\hspace{0.2cm}
\subfloat[]{\includegraphics[width=0.45\textwidth]{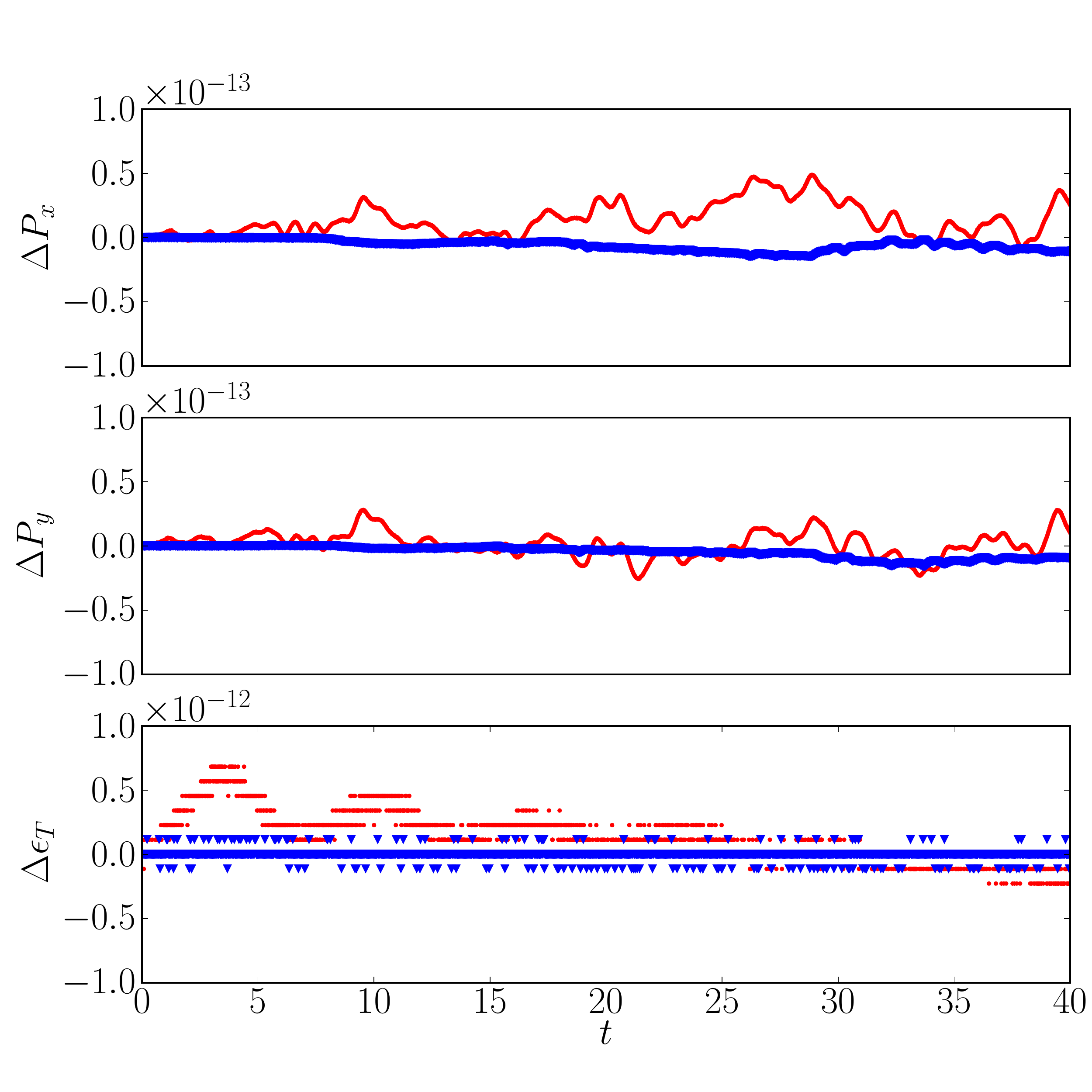}}
\caption{\label{iawfig} a) Ion velocity moment perturbation amplitude from the electrostatic ion acoustic wave test on non-orthogonal sinusoidal (red) and uniform Cartesian (blue) meshes. The black dashed line shows the theoretical damping rate with slope $\gamma = -0.0744636$. b) Numerical conservation errors. Top: $\boldsymbol{\hat{x}}$-component of linear momentum. Middle: $\boldsymbol{\hat{y}}$-component of linear momentum. Bottom: Total (ion + electron kinetic) energy. These errors are comparable to the non-linear tolerance of $10^{-12}$, as expected. }
\end{center}
\end{figure}

Figure~\ref{iawfig} (left) shows the amplitude of the ion velocity moment vs time for the sinusoidal mesh (red) of Eqs.~(\ref{sinmesheq1},~\ref{sinmesheq2}), as well as a uniform Cartesian mesh with the same resolution (blue). The amplitude of the wave decreases exponentially with time due to kinetic ion Landau damping. The theoretical damping rate is found from the dispersion relation $Z^\prime (\xi) = 2\tau$, where $Z$ is the plasma dispersion function and $\xi = (\omega + i \gamma)/k\sqrt{2}v_{Ti0}$ is the normalized complex frequency. For our initial conditions, the theoretical value is $\gamma = -0.0744636$, which is shown as the black dashed line on Fig.~\ref{iawfig} (left). There is good agreement for both the Cartesian and curvilinear mesh cases with this theoretical damping rate until the noise due to a finite number of particles overcomes the signal. 

The conservation errors are shown in Fig.~\ref{iawfig} (right) for both the Cartesian mesh (blue) and curvilinear mesh (red) simulations. The momenta and energy are conserved to better than $10^{-13}$ and $10^{-12}$, respectively, which is on the level of the non-linear tolerance ($10^{-12}$) used.  This provides numerical verification of the result of Section~\ref{electrostaticmomentum}, that, for the electrostatic limit, the total linear momentum and energy is conserved for general curvilinear mapped meshes.

\subsection{Spatial convergence for 2D whistler wave propagation}\label{whistler}

In the second numerical example, we simulate the right-hand polarized Alfv\'{e}n-whistler wave in a cold and uniform background plasma. This wave has an exact analytical solution to the cold-plasma hybrid model for arbitrary perturbation amplitude, and therefore can be used to perform a formal grid convergence study. All simulations are performed in 2D, with the initial background plasma specified by $Z_i=M_i=N_i=1$, $T_i=T_e=0$, and the magnetic field oriented diagonally in physical space as $\boldsymbol{B}/B_0 = (\boldsymbol{\hat{x}} + \boldsymbol{\hat{y}})/\sqrt{2}$. 

The eigenmode of the wave is perturbed as
\begin{equation}\boldsymbol{\delta B} = \boldsymbol{\nabla} \times \boldsymbol{\delta A} = 10^{-3}\boldsymbol{\nabla} \times \left( \frac{\cos{(\boldsymbol{k}\cdot \boldsymbol{x})}}{k_x}\boldsymbol{\hat{y}} - \frac{\sin{(\boldsymbol{k}\cdot \boldsymbol{x})}}{k}\boldsymbol{\hat{z}}\right), \end{equation}
\begin{equation}\frac{\boldsymbol{\delta u}}{v_{A0}} = - \frac{\boldsymbol{\delta B}}{B_0} \frac{v_{A0}k}{\omega} ,\end{equation}
where $k_xd_i=k_yd_i=2\pi/16$ and $kd_i=\sqrt{k_x^2+k_y^2} \approx 0.56$, and $\omega = v_{A0}k \left(\tfrac{1}{2}d_ik + \sqrt{1 + \tfrac{1}{4} d_i^2 k^2}\right)$. This intermediate value of $k d_i$ is chosen to simultaneously test both the grid-based electron and particle-based ion discretizations. These simulations use $400$ particles/cell initialized with the quasi-quiet start, and apply one pass of conservative binomial smoothing (\ref{conservativesmoothing}). The timestep used is $\Delta t \Omega_{ci}=0.0125$.

\begin{figure}
\begin{center}
\includegraphics[width=0.85\textwidth]{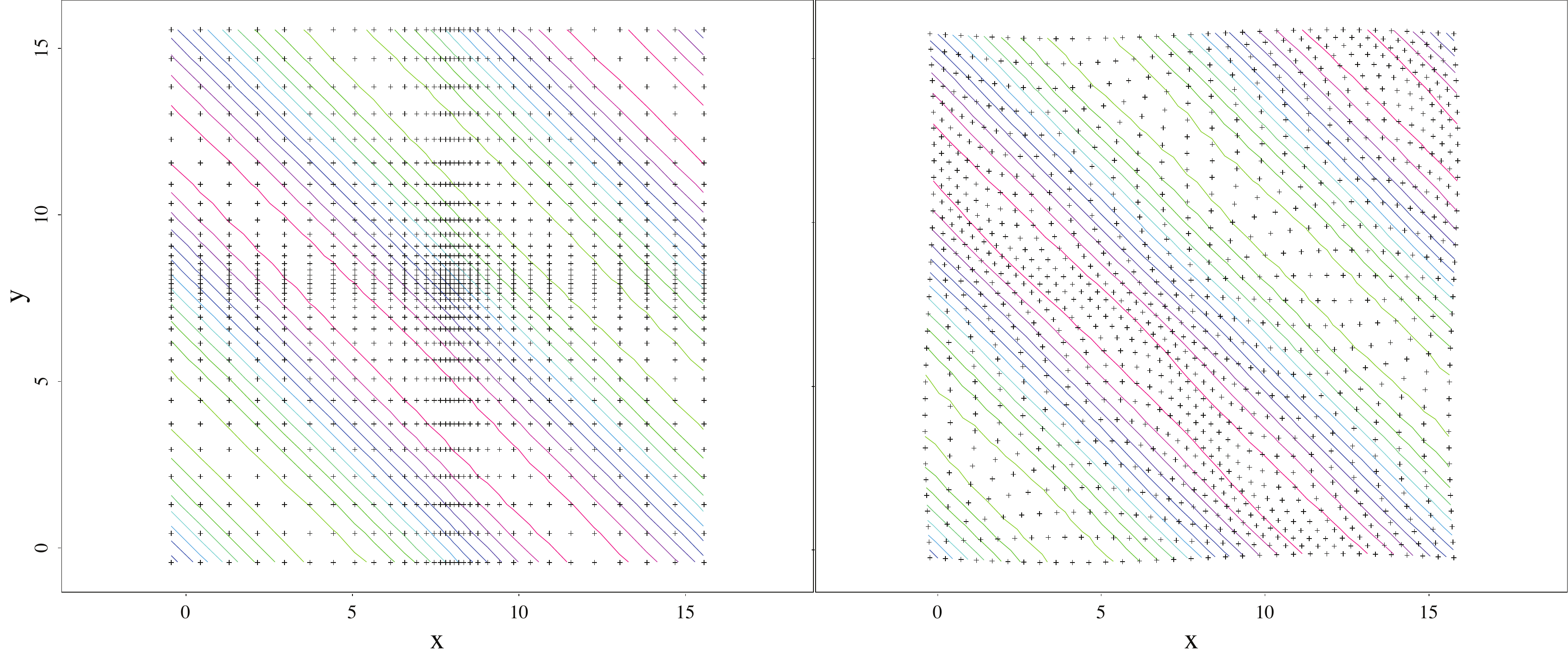}
\caption{\label{whistlergrids}Meshes used for Alfv\'{e}n-whistler wave propagation problem. The background magnetic field and direction of propagation are along the diagonal ($45^\circ$) of the 2D domain. The colored lines show contours of the magnetic vector potential component $A_z$, and the black $+$ symbols show the centers of the spatial cells. Left: Cartesian tensor-packed mesh with packing factor of 4 in the $x$ and $y$ directions. Right: Non-orthogonal sinusoidal mesh of Eqs.~(\ref{sinmesheq1}-\ref{sinmesheq2}) with distortion factor $\sigma=0.4$.}
\end{center}
\end{figure}

Figure~\ref{whistlergrids} shows the two types of mesh that are used for this numerical test. The left panel shows a Cartesian tensor-packed mesh of the kind discussed in Section~\ref{planargrids} that is packed by a factor of $4\times$ in the $x$ and $y$ directions, and the right panel shows a non-orthogonal sinusoidal mesh of Eqs.~(\ref{sinmesheq1}, \ref{sinmesheq2}) with distortion factor $\sigma=0.4$.

\begin{figure}
\begin{center}
\includegraphics[width=0.55\textwidth]{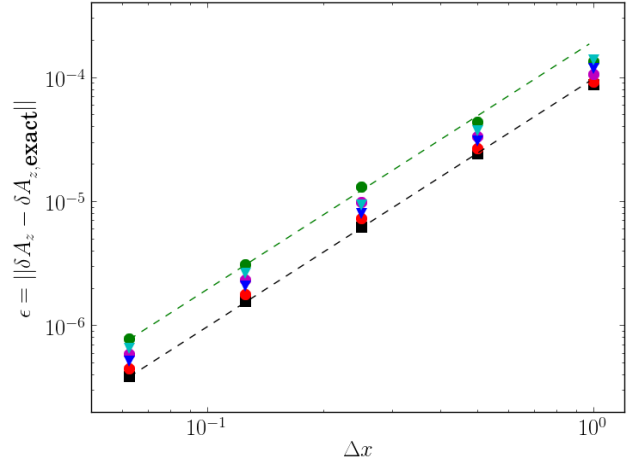}
\caption{\label{whistlerconvergence}Spatial grid convergence test for whistler-wave 2D propagation problem. The uniform-mesh case is given by black squares. Circles are from the sinusoidal grid runs with $\sigma = 0.5$ (red), $\sigma=1$ (magenta), and $\sigma=1.5$ (green). Triangles are from the stretched planar mesh runs with stretch factors of $2\times$ (blue), and $4\times$ (cyan). The black and green dashed lines show ideal $(\Delta x)^2$ convergence.}
\end{center}
\end{figure}

Figure~\ref{whistlerconvergence} shows a formal spatial convergence study of the numerical scheme for the Cartesian tensor-packed mesh as triangles (packing factor of $2\times$ in blue, and $4\times$ in cyan), the sinusoidal mesh as circles ($\sigma = 0.5$ in red, $\sigma = 1$ in magenta, and $\sigma=1.5$ in green), as well as a uniform Cartesian mesh (black squares) for comparison. The black and green dashed lines are shown for reference, both with $\epsilon \propto (\Delta x)^2$. This order of convergence holds for all of the different spatial meshes.

\begin{figure}
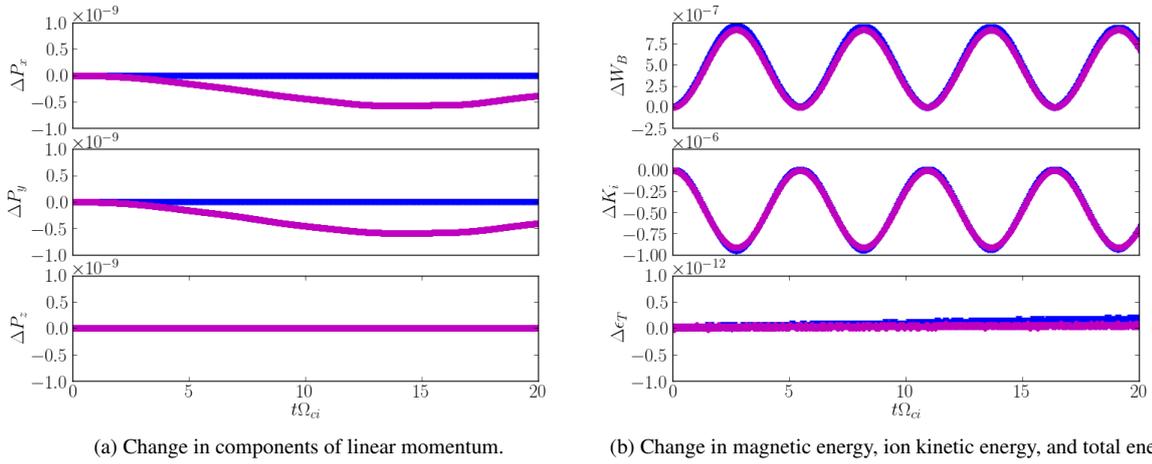

\begin{center}
\subfloat[Change in components of linear momentum.]{\includegraphics[trim={0cm 0cm 0cm 0cm},clip,width=0.48\textwidth]{momentum-magenta.pdf}}
\subfloat[Change in magnetic energy, ion kinetic energy, and total energy.]{\includegraphics[trim={0cm 0cm 0cm 0cm},clip,width=0.48\textwidth]{energy-magenta.pdf}}
\caption{\label{momentumenergywhistler} Conservation errors in momentum and total energy for Alfv\'{e}n-whistler wave problem. The blue curve shows results for a Cartesian tensor-packed mesh with packing factor of $2\times$, and the magenta curve is from the non-orthogonal sinusoidal mesh simulation with distortion factor $\sigma = 1$. }
\end{center}
\end{figure}

Figure~\ref{momentumenergywhistler} (left) shows the three components of the linear momentum error, $\boldsymbol{\Delta P} = \boldsymbol{P}^n - \boldsymbol{P}^0$, for the Cartesian tensor-packed mesh with a packing factor of $2\times$ (blue) and the sinusoidal mesh with $\sigma=1$ (magenta). Both simulations have $n_\xi=n_\eta = 128$, with other parameters the same as for Fig.~\ref{whistlerconvergence}. In agreement with Section~\ref{planargrids}, the Cartesian tensor-packed meshes show very small momentum errors $(\textrm{max}\left[|\boldsymbol{\Delta P}|\right] \approx 10^{-13}$) for a non-linear tolerance of $10^{-12}$, whereas the sinusoidal mesh has a larger momentum error, $\textrm{max} \left[| \boldsymbol{\Delta P }|\right] \approx 6\times 10^{-10}$.

Figure~\ref{momentumenergywhistler} (right) shows the energies from the same two simulations. The top two panels show the variation in the magnetic energy ($\Delta W_B = W_B^{n} - W_B^0$) and ion kinetic energy ($\Delta K_i = K_i^n - K_i^0$), respectively, and the bottom panel shows the sum of the two. The total energy errors for both cases are comparable, with values $\textrm{max}\left[\Delta \epsilon_T\right] \approx 10^{-13}$. We have verified that this error scales with the chosen non-linear tolerance of the iterative solver, as expected.

\subsection{Non-linear magnetic island coalescence problem}\label{islandcoal}

The third numerical example is the electromagnetic island coalescence problem, which is a non-linear and multi-scale magnetic reconnection challenge problem~\cite{finn1977,pritchett1992coalescence,karimabadi11,stanier15prl,ng15,stanier17,makwana2018,allmann2018}. Recent simulation studies have highlighted the key role that the ion kinetic physics has in determining the rate of magnetic reconnection, and the global motions of the interacting magnetic islands~\cite{stanier15prl}. 

\begin{figure}
\begin{center}
\subfloat{\includegraphics[trim={0cm 0.2cm 1cm 3.2cm},clip,width=0.85\textwidth]{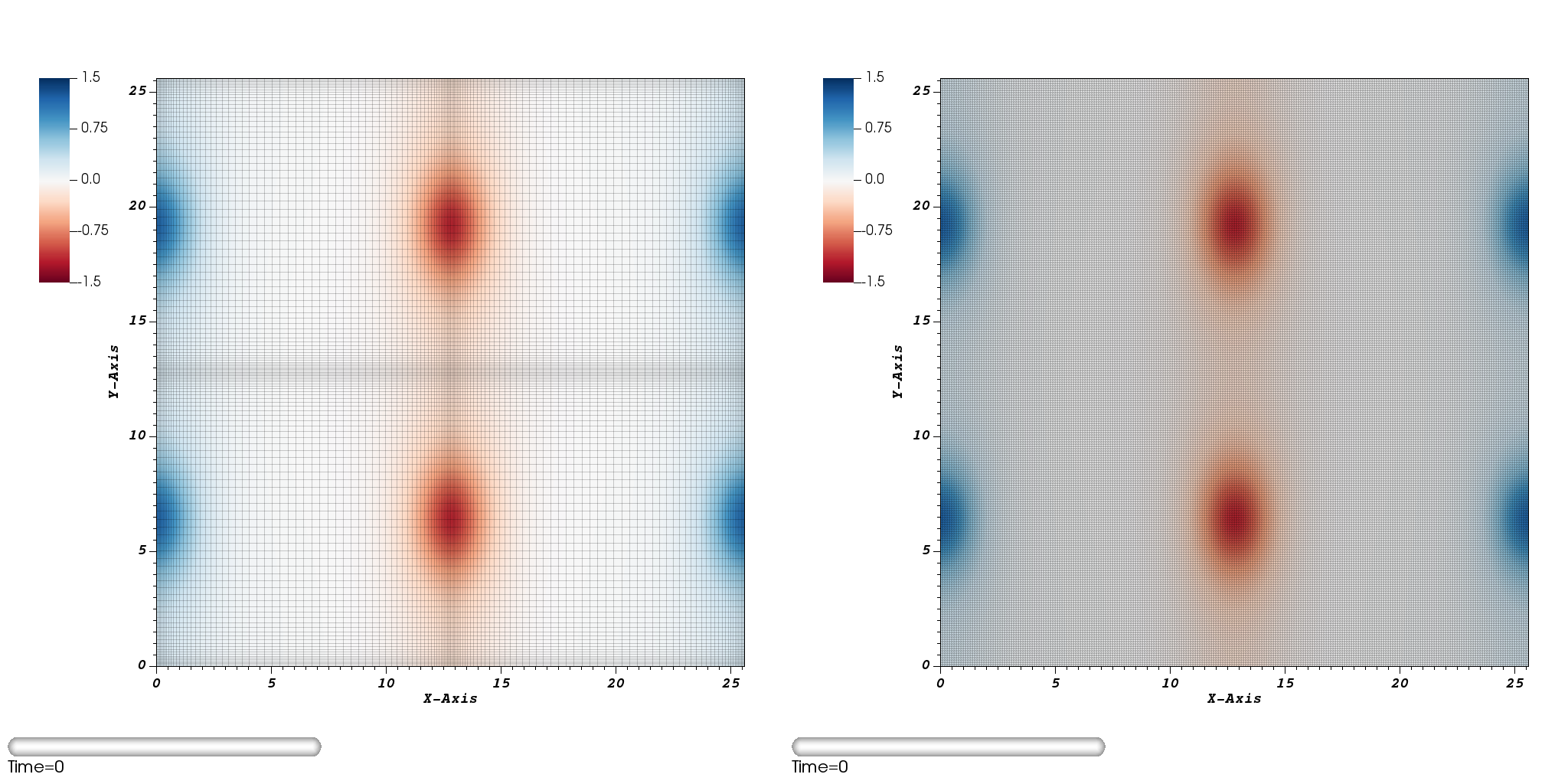}}\\
\subfloat{\includegraphics[trim={0cm 0.2cm 1cm 3.2cm},clip,width=0.85\textwidth]{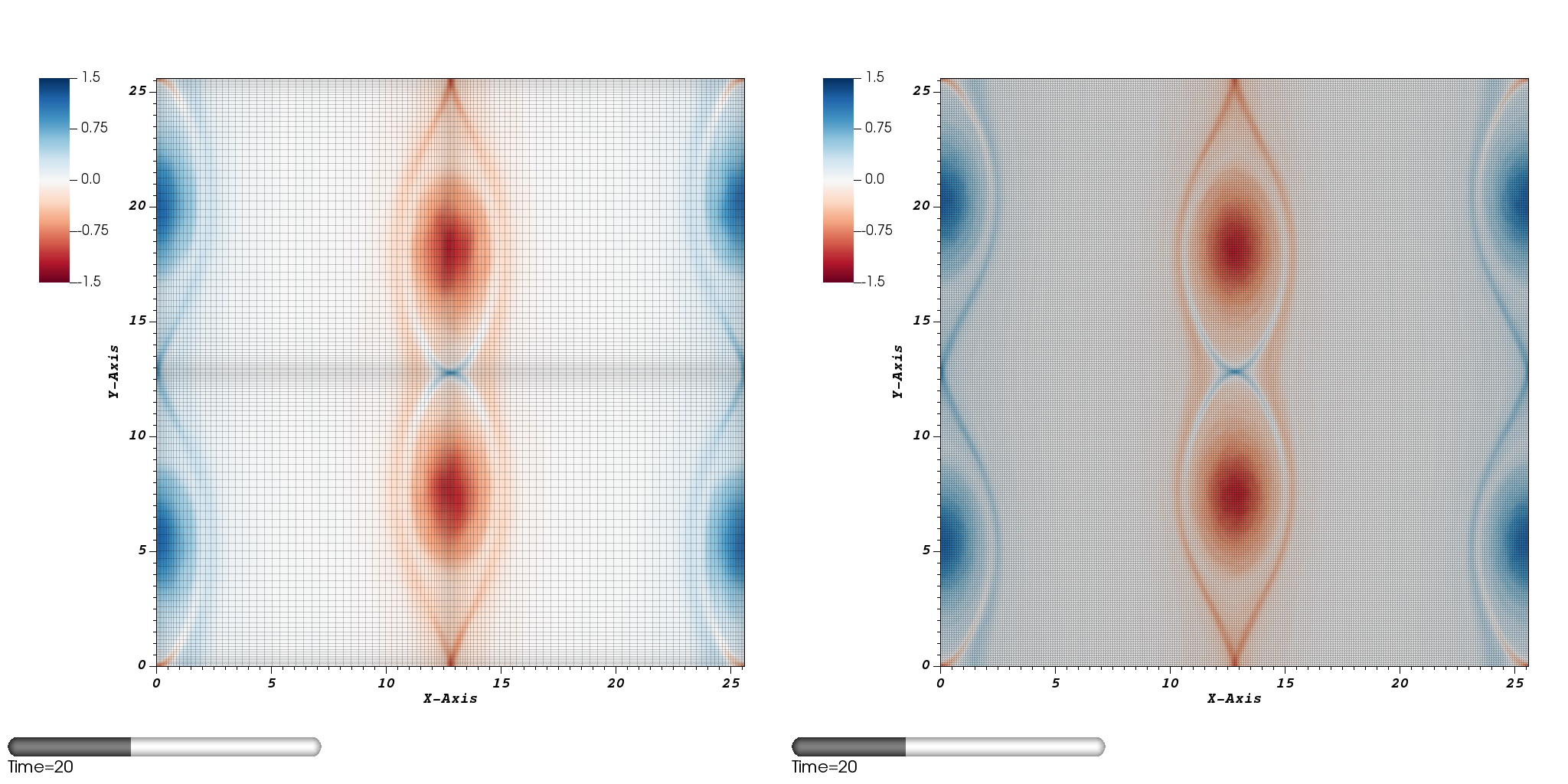}}\\
\subfloat{\includegraphics[trim={0cm 0.2cm 1cm 3.2cm},clip,width=0.85\textwidth]{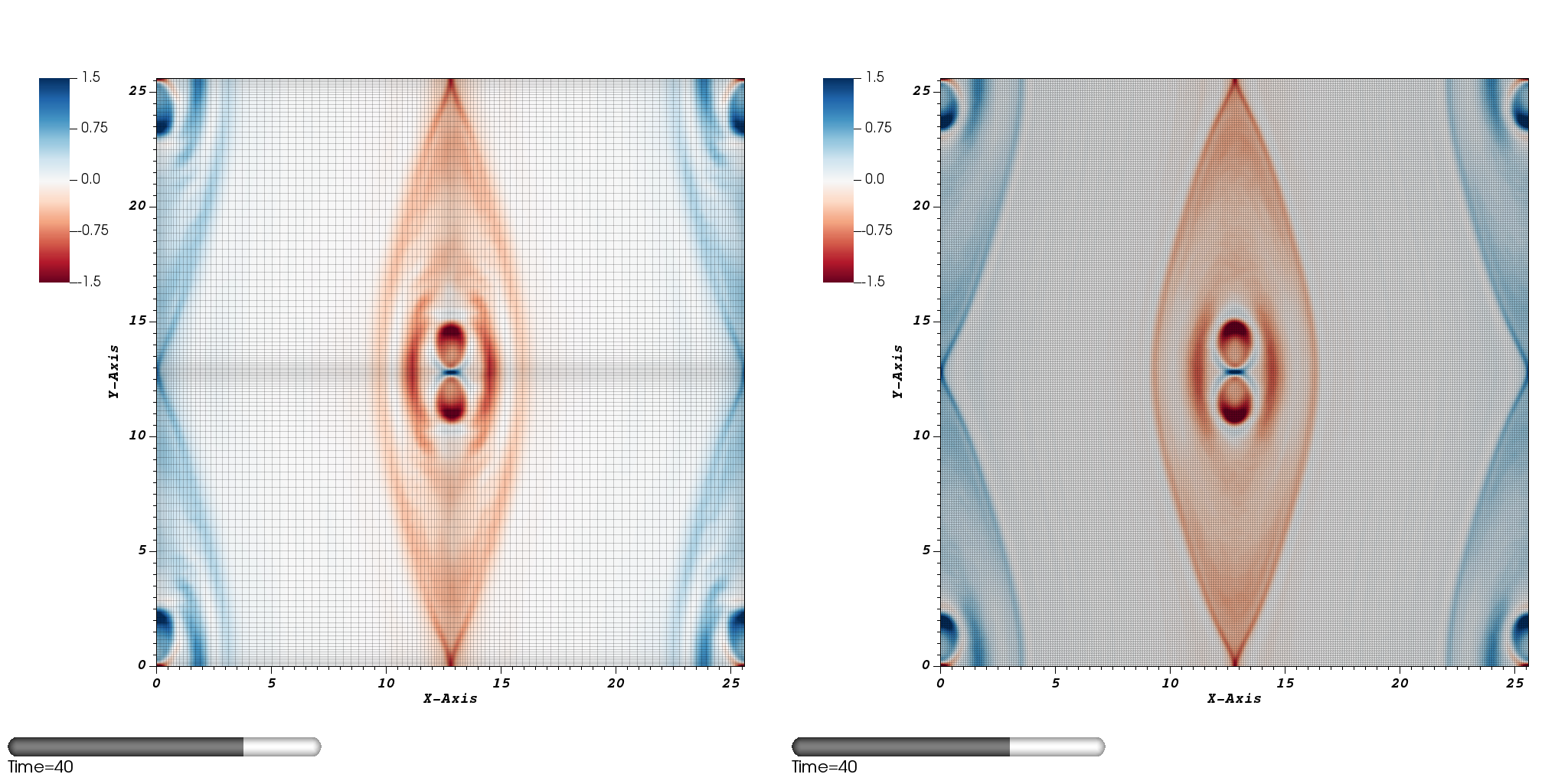}}
\caption{\label{jz-mesh-island}Pseudocolor plot of out-of-plane current density $j_z$ from the island coalescence problem with $4$ initial islands ($2$ are shown at the periodic boundary in $x$). Left column: Cartesian tensor-packed mesh (shown as black lines) with $n_\xi=n_\eta=128$ and stretch factor of $4\times$ in $x$ and $y$ directions. Fine resolution is placed along the simulation boundaries and in the center of the domain. Right column: Uniform Cartesian mesh with $n_x=n_y=512$. Both simulations use $n_{ppc}=400$, so the tensor-packed mesh case has $16\times$ fewer particles. }
\end{center}
\end{figure}

We assume periodic boundary conditions at $x=0, L_x$, and $y=0, L_y$, such that there are four magnetic islands in total within the domain - two islands cross the periodic boundary at $x=0$ and $x=L_x$ as shown in Figure~\ref{jz-mesh-island}. The initial conditions given below are in an unstable ideal-magnetohydrodynamic equilibrium state, where magnetic islands with the same sign of the out-of-plane current density $j_z$ attract each-other due to the Lorentz force.

\begin{figure}
\begin{center}
\subfloat[Change in components of linear momentum.]{\includegraphics[trim={1.8cm 1cm 1.8cm 1cm},clip,width=0.48\textwidth]{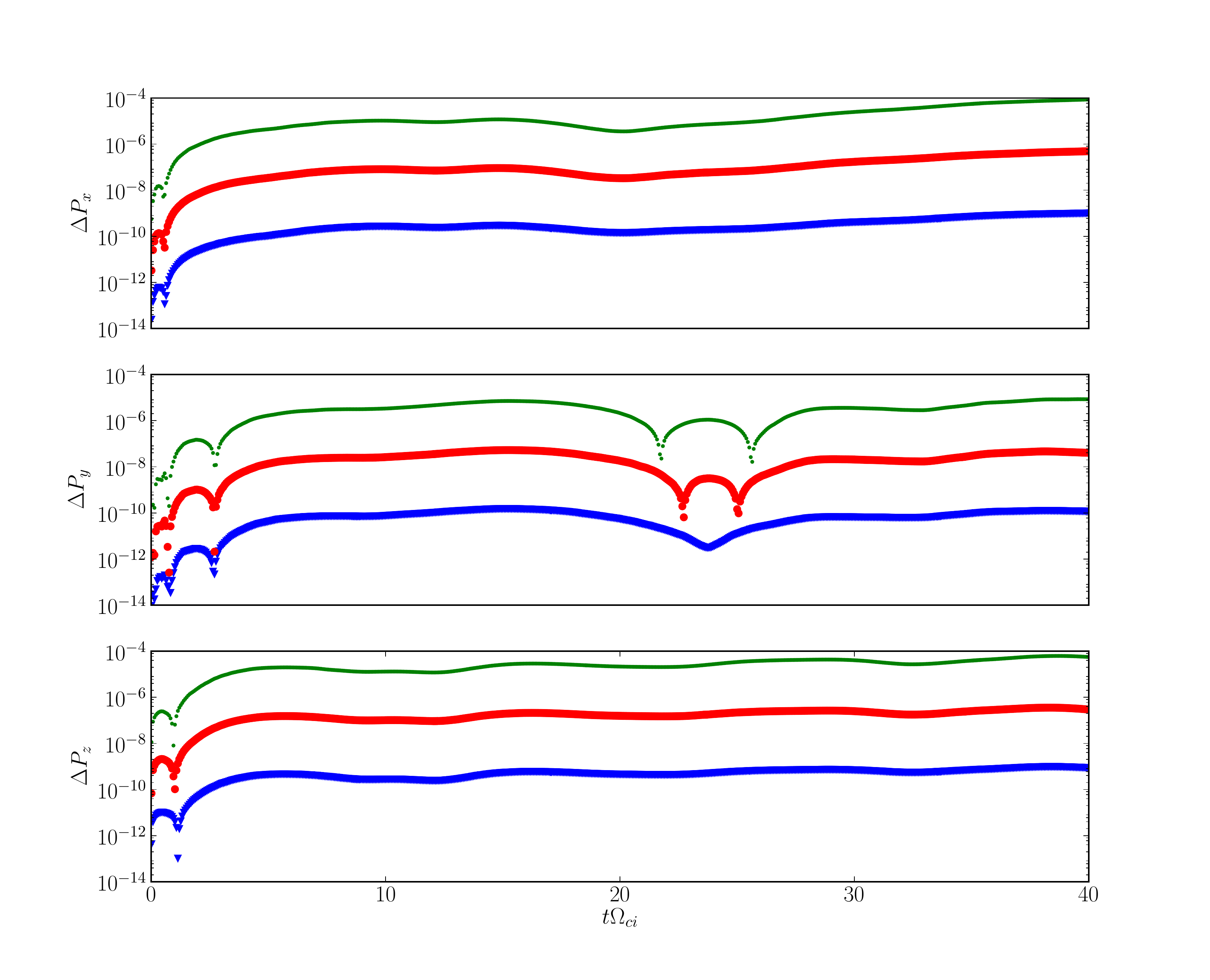}}
\subfloat[Change in magnetic energy, electron kinetic energy, ion kinetic energy, and total energy.]{\includegraphics[trim={1.8cm 1cm 1.8cm 1cm},clip,width=0.48\textwidth]{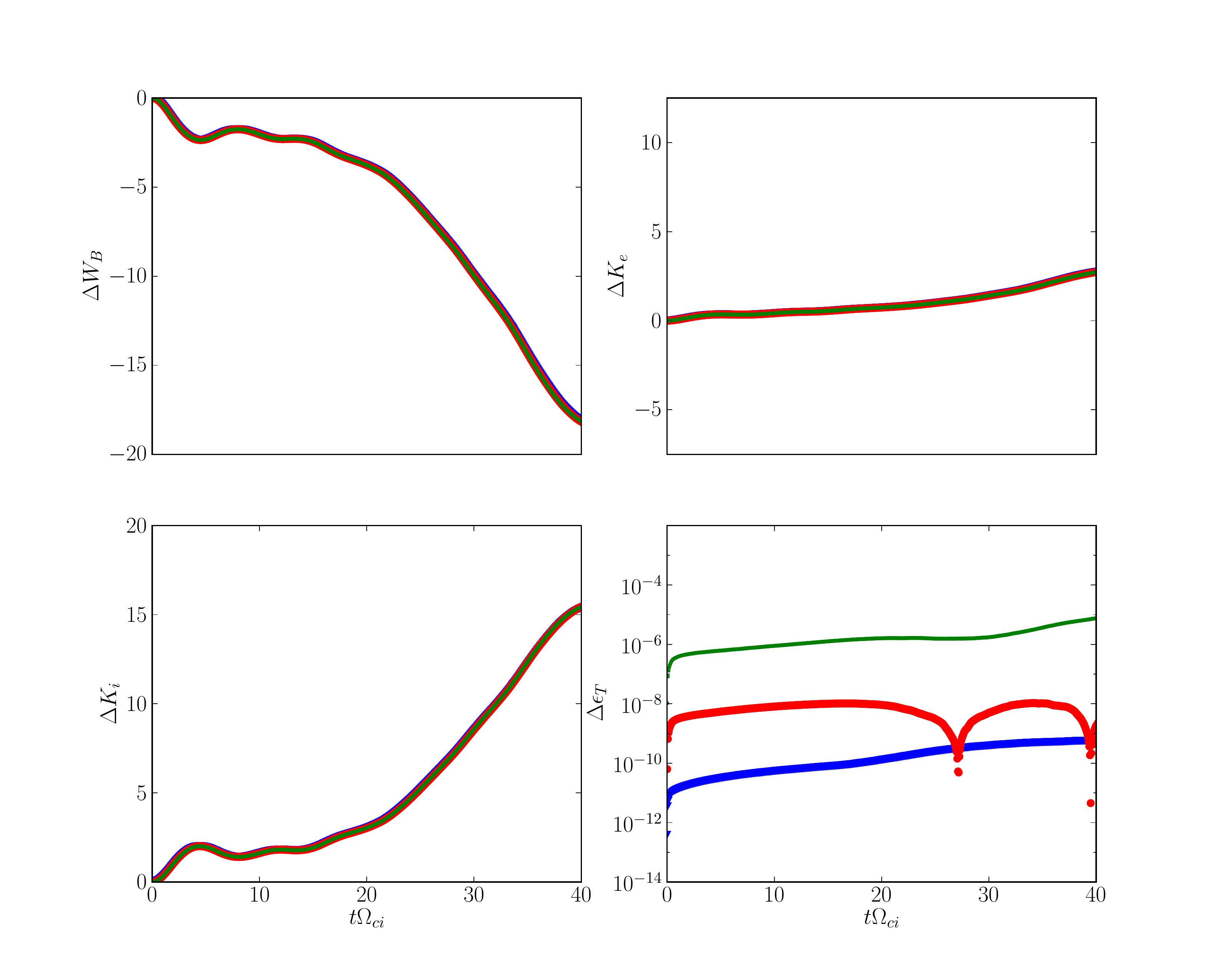}}
\caption{\label{islandcoal-momen} Conservation errors in momentum and total energy for the tensor-packed mesh island coalescence simulation at three values of the relative non-linear tolerance (exact conservation is expected for arbitrary small tolerance). }
\end{center}
\end{figure}

A single ion species is used with $Z_i=M_i=1$. The ions are initialized with a Maxwellian distribution function with initial density profile~\cite{fadeev1965} 
\begin{equation}N_i=0.2 + \frac{1-\alpha^2}{(\cosh{(x/\lambda)} + \alpha\cos{(4\pi y/L_y)})^2} + \frac{1-\alpha^2}{(\cosh{((x-L_x/2)/\lambda)} + \alpha\cos{(4\pi y/L_y)})^2} +\frac{1-\alpha^2}{(\cosh{((x-L_x)/\lambda)} + \alpha\cos{(4\pi y/L_y)})^2},\end{equation}
where $\alpha=0.4$, $\lambda=2d_i$, and $L_x=L_y=25.6 d_i$. The magnetic vector potential is given by  
\begin{eqnarray}A_z =& -\lambda \{\ln{[\cosh{(x/\lambda)} + \alpha \cos{(4\pi y/L_y)}]}+\ln{[\cosh{((x-L_x)/\lambda)} + \alpha \cos{(4\pi y/L_y)}]} \nonumber \\
&-\ln{[\cosh{((x-0.5L_x)/\lambda)} + \alpha \cos{(4\pi y/L_y)}]}\}.\end{eqnarray}
The ion plasma beta is $\beta_i = 5/6$ with an ion-to-electron temperature ratio $\tau = 5$ (total ion $+$ electron $\beta = 1$). The plasma resistivity used is $\eta = 5\times 10^{-4}$, and the heat conductivity is $\kappa = 0$. Here, we also include a higher-order \textit{hyper-resistive} contribution to the resistive friction term such that $F_{ie} = \eta \boldsymbol{j} - \eta_H \nabla^2 \boldsymbol{j}$ with $\eta_H=2\times 10^{-4}$. This is used to set a dissipation scale for the thin current sheets that form in Fig.~\ref{jz-mesh-island}, otherwise these can shrink to the mesh-scale~\cite{stanier15b}. We note that the hyper-resistivity is not needed for numerical stability in this case, and the runs shown in Fig.~\ref{islandcoal-momen} do not include the hyper-resistivity. To break the initial symmetry, we apply a sinusoidal perturbation with amplitude $\delta A_z = 0.1$, which causes the negative current islands (red) to move towards the midplane ($y=L_y/2$), and the positive current islands (blue) to move towards the outer boundaries at $y=0$ and $y=L_y$. Magnetic reconnection occurs at each of the magnetic $X$-points, located at $(x,y) = (0,0)$, $(0,L_y/2)$, $(L_x/2,0)$, $(L_x/2,L_y/2)$, and the matching points on periodic boundaries $x=L_x$ and $y=L_y$.

Figure~\ref{jz-mesh-island} shows a comparison between a simulation with a Cartesian tensor-packed mesh with resolution $n_\xi=n_\eta=128$ (left column) and a $n_x=n_y=512$ resolution uniform Cartesian mesh (right column). Both cases use $400$ particles/cell which are initialized with a quasi-quiet start, and use quadratic-spline shape functions with two passes of conservative smoothing (\ref{conservativesmoothing}). The timestep used is $\Delta t\Omega_{ci} = 0.0125$. The map used in the tensor-packed mesh is chosen to give the same resolution at the magnetic X-points as for the uniform mesh case. However, as the same number of average particles per cell is used, the tensor-packed mesh case uses $16\times$ fewer particles with corresponding reductions in both CPU time and memory usage. Despite this, the rate of coalescence of the magnetic islands is very similar between the two simulations as shown by the similar size of the islands at $t\Omega_{ci} = 40$ in the bottom row.

Figure~\ref{islandcoal-momen} shows the momenta and energies from the same tensor-packed mesh simulation of Fig.~\ref{jz-mesh-island} (left) but with zero hyper-resistivity ($\eta_H=0$) for three different values of the relative non-linear tolerance: $10^{-4}$ (green), $10^{-6}$ (red), and $10^{-8}$ (blue). For these tensor-packed Cartesian meshes, both the momentum and total energy errors are reduced for tighter values of the non-linear tolerance, as expected. 


\subsection{Kink ($m=1$) mode in helical geometry}\label{m1helical}

\begin{figure}
\begin{center}
\includegraphics[width=0.8\textwidth]{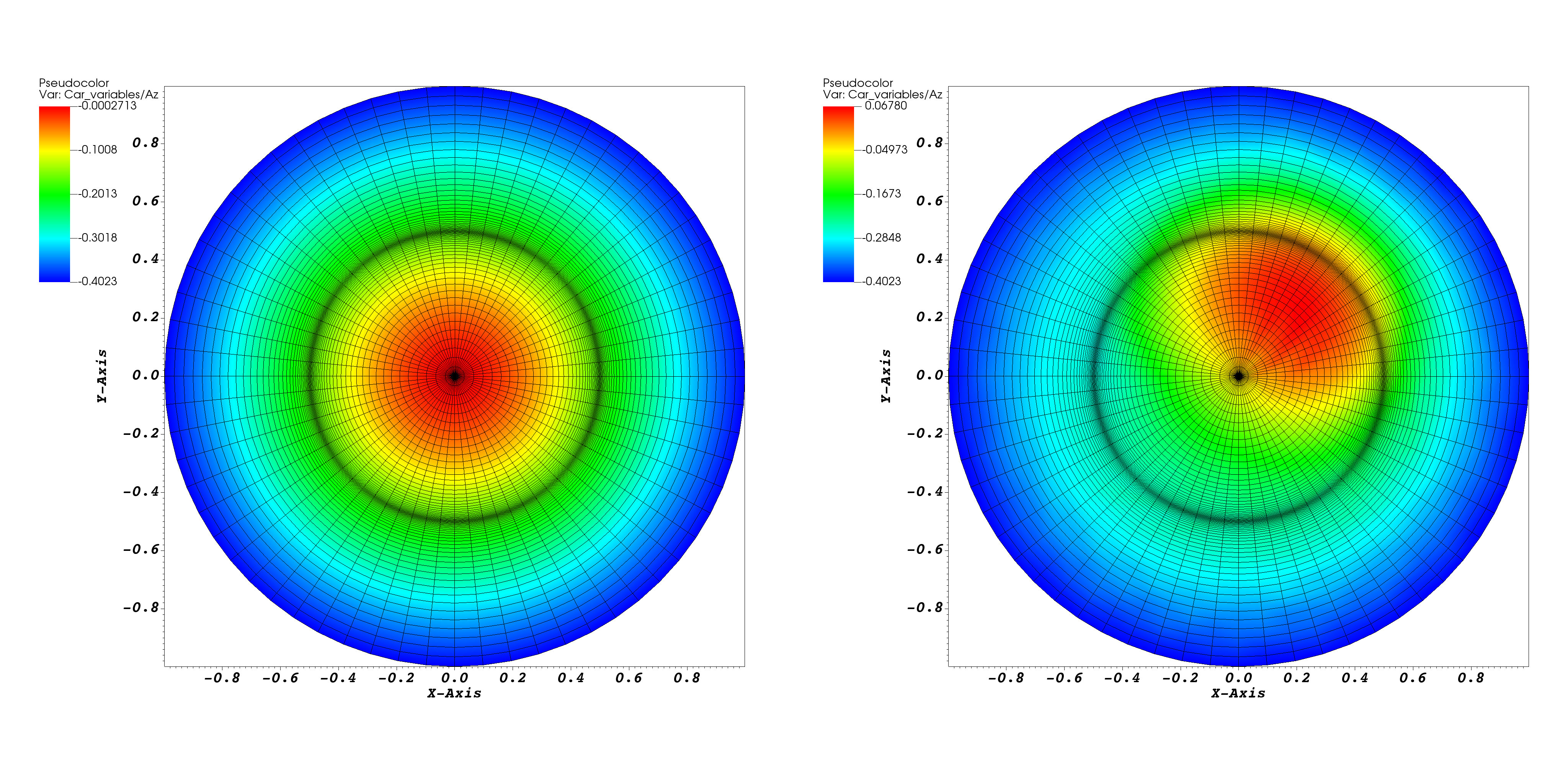}
\caption{\label{m1azplots}Magnetic vector potential $A_z$ at initial ($t=0$) and final time ($t=9.5\tau_0$) from the hybrid-PIC simulation of the $m=1$ mode in 2D helical geometry. Also shown is the 2D helical mesh which is packed in the radial direction at the location of the initial rational surface. }
\end{center}
\end{figure}

In the final numerical example, we illustrate the algorithm for a more challenging geometry with non-periodic boundary conditions. The $m=n=1$ mode of a cylindrical plasma column is simulated on a 2D non-orthogonal helically symmetric mesh with an underlying map given by $x=r\cos{\tau}$, $y=r\sin{\tau}$. Here, $\tau = (\theta - kz)/m$ is the helical angle, $\theta$ the poloidal angle, $z$ the axial position, $k=n/R$ the axial wavenumber, and $R$ is the major radius of the cylinder which is a topological torus due to periodic boundary conditions in the poloidal and axial directions. In addition, we convolve this map with a packing function that is non-uniform in the radial direction such that resolution is finer at the rational surface ($\Delta r_{\textrm{min}} = 0.002$ at $r=0.5$) and decreased in the cells adjacent to the singular point ($r\approx 0$). The latter is helpful to alleviate high levels of statistical noise that otherwise can occur in these cells due to their extremely small cell volumes and thus low numbers of particles per cell. The mesh used for the simulation is shown in Fig.~\ref{m1azplots}. The spatial mesh has $[n_r, n_\tau] = [64, 64]$ spatial points with an average of $400$ macro-particles per cell, but these are distributed non-uniformly to place more particles into regions of low cell volumes that are more susceptible to noise as the simulation progresses. The boundary conditions are a singular-point boundary at $r=0$ and a conducting outer boundary at $r=1$, which are described in~\ref{sec:boundaryconditions}. We use a tensor product of zero and second-order shape functions~(\ref{sec:boundaryconditions}) and apply two passes of conservative smoothing (\ref{conservativesmoothing}). The timestep used is $\Delta t = 0.01$ and the non-linear tolerance is $\epsilon_t = 10^{-3}$.

We use a single ion species with $Z_i=M_i=1$. The initial profiles are 1D cylindrically symmetric profiles, which form an exact Vlasov equilibrium~\cite{bennett1934}. Macroscopically, the columns magnetic pinch force is balanced by a thermal pressure gradient that points radially outwards. The plasma density and poloidal magnetic field profiles are given by
\begin{equation}\label{m1density}N_i = \frac{1}{\left(1+(r/\lambda)^2\right)^2},\end{equation}
\begin{equation}\label{m1btheta}B_\theta/B_0 = \frac{\left(r/\lambda\right)}{\left(1 + (r/\lambda)^2\right)}\end{equation}
with $B_z/B_0=1$, such that the helical field $B_\tau/B_0 = m (B_\theta/B_0r) + k$. Initial equilibrium requires that $\beta_T = 2 n_0 (T_{i0} + T_{e0})/B_0^2 = 1$. The physical parameters used are an ion-to-electron temperature ratio $\tau = T_i/T_e = 0$ (chosen to compare with a cold-ion Hall-MHD model, see below), resistivity $\eta = 2\times 10^{-3}$ and heat conductivity $\kappa = 2\times 10^{-3}$. We also note that the normalization for this simulation differs from Sec.~\ref{sec:normalize}, as we set $L_0$ as the minor radius with $d_i/L_0 = 0.1$. Similarly, the time-scale is normalized by the Alfven time $\tau_0 = L_0/v_{A0}$.

We apply a small perturbation to the radial ion bulk velocity moment
\begin{equation}\label{m1pert}\delta u_r = 10^{-3}\sin^3{(\pi r)}\cos{(\theta)}\end{equation}
to displace the column from equilibrium.

In order to sample the initial density profile from Eq.~(\ref{m1density}) and velocity moment perturbation from Eq.~(\ref{m1pert}) to a high degree of accuracy for the non-uniformly distributed macro-particle profile, we compute the individual particle weights ($w_p$) and velocities ($\boldsymbol{v}_p^0$) by inverting a mass-matrix~\cite{chen2021}.  These particles are initialized with a quasi-quiet start.

Figure~\ref{m1azplots} shows the component of the magnetic vector potential $A_z$ (color scale) along with the spatial mesh (black lines) for the initial and final time of the simulation. At the final time, the plasma column shows a characteristic helical kink-like displacement from the origin. This $m=1$ mode is driven by the gradients in the current density and plasma pressure.

\begin{figure}
\begin{center}
\subfloat[Radial profile of $u_\theta$ in linear growth phase]{\includegraphics[trim={0cm -3.8cm 0cm -3.8cm},clip,width=0.48\textwidth,valign=c]{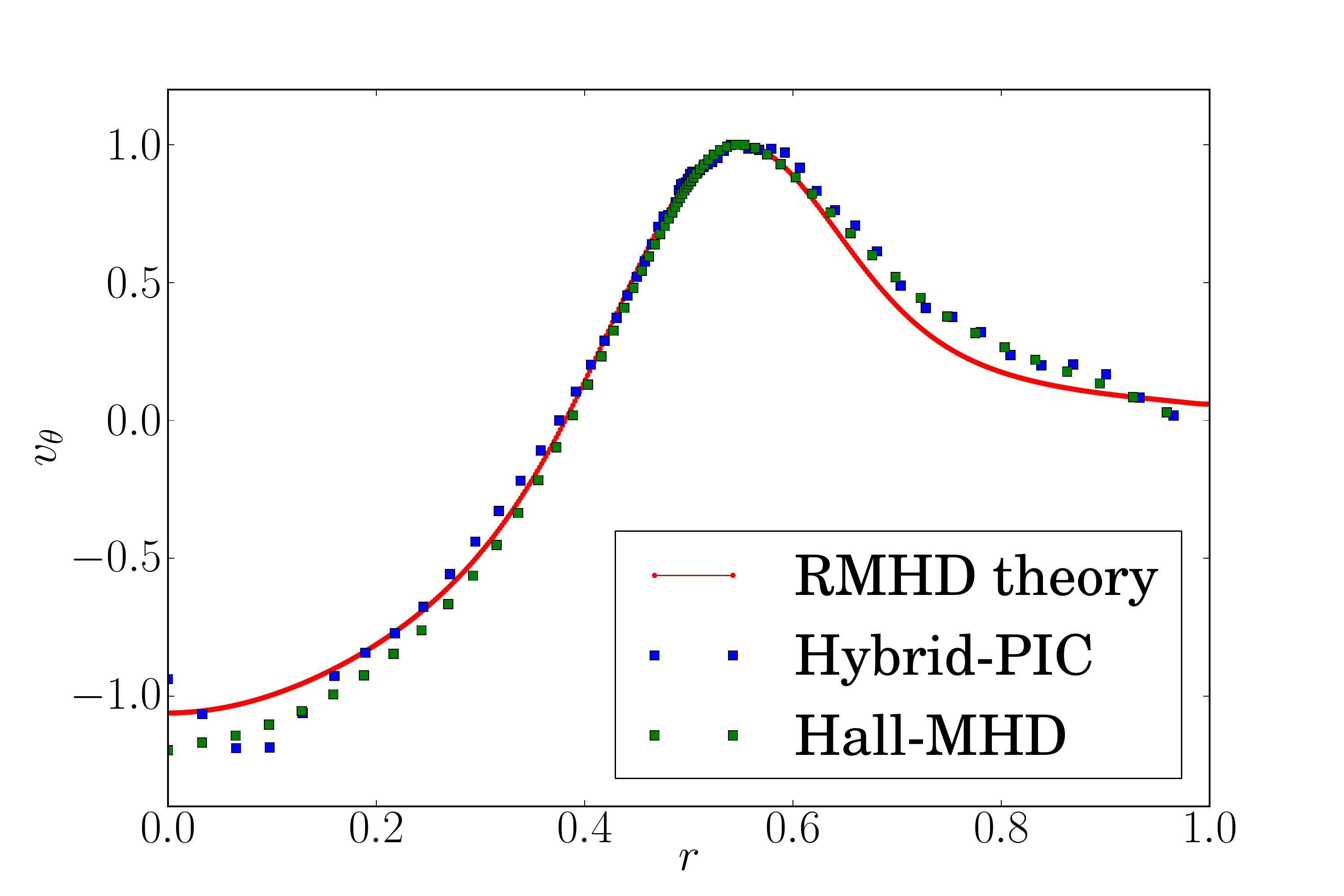}}
\subfloat[Mode amplitudes of density (top) and magnetic vector potential (bottom) perturbations]{\includegraphics[trim={0cm 0cm 0cm 0cm},clip,width=0.45\textwidth,valign=c]{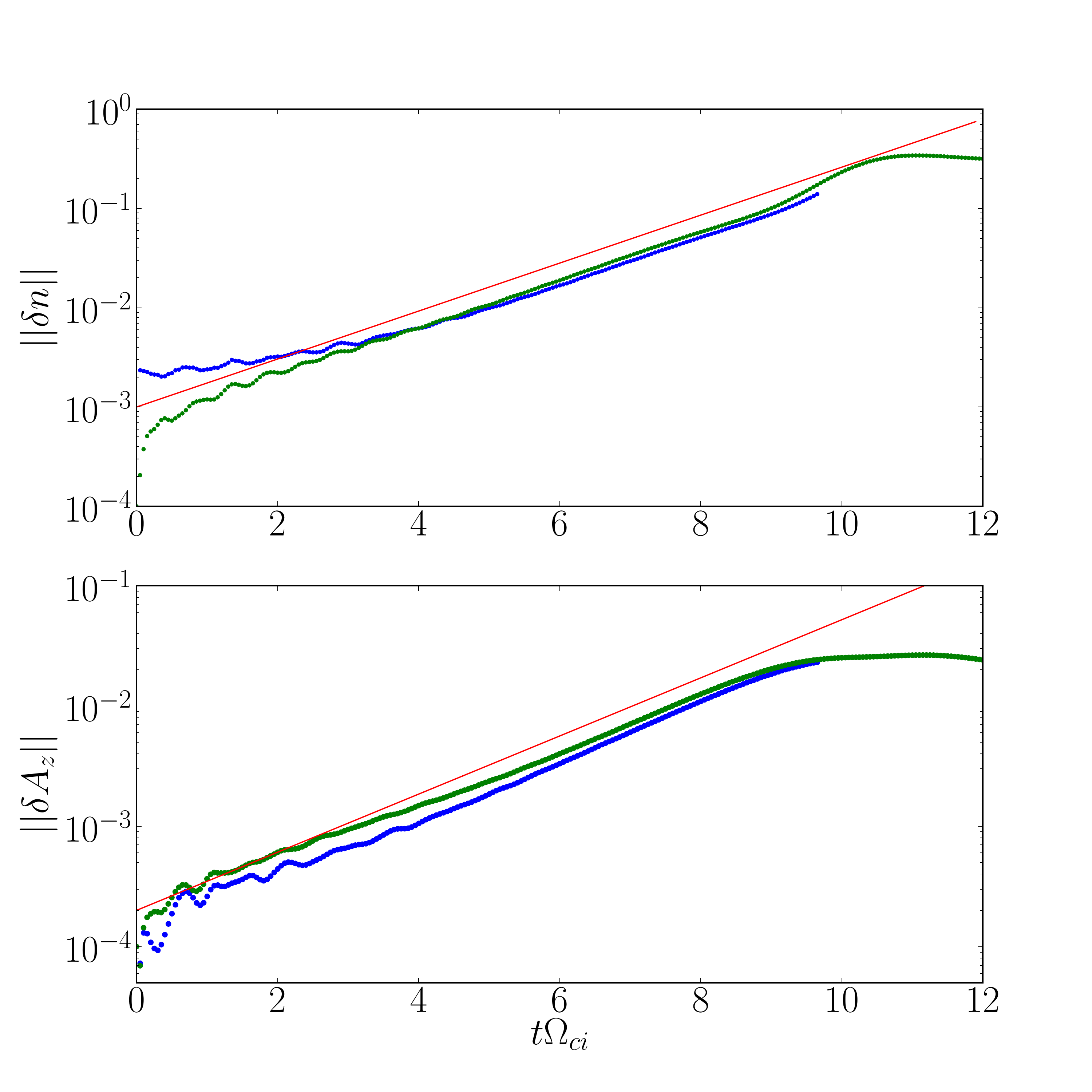}}
\caption{\label{m1eigenfunction}Eigenfunction of $m=1$ mode in helical geometry simulation. Green: Simulation result from non-linear Hall-MHD simulation. Blue: Simulation from cold ion hybrid simulation. The linear theory result (red) is computed for resistive-MHD (RMHD), which is in reasonable agreement since the inertial layer width is larger than the ion kinetic scales for these parameters. }
\end{center}
\end{figure}

Figure~\ref{m1eigenfunction} (left) shows a comparison of the radial profile of the poloidal bulk velocity $u_\theta$ from the simulation, along with the linear eigenfunction of the $m=1$ mode. This eigenfunction and the theoretical linear growth rate (red lines in the right panel of Fig.~\ref{m1eigenfunction}) are computed numerically by solving the linearized resistive-MHD equations in cylindrical geometry using the equilbrium specified above. Although the resistive-MHD theory misses some important diamagnetic and finite gyro-radius physics -- which gives a real frequency of the mode and rotation observed in Fig.~\ref{m1azplots} (right) -- our choice of parameters ensures that these differences remain small. In particular, we pick initial conditions which have a large MHD drive such that the inertial layer thickness $\Delta_{in}$ is larger than the ion kinetic scales~\cite{mishchenko12}, which allows comparison with resistive-MHD theory.  We also verify the results against a non-linear Hall-MHD simulation with the same initial set-up (shown in green). In the limit of cold ions, the hybrid-PIC and Hall-MHD models are physically equivalent but there are significant algorithmic differences, namely that Hall-MHD model uses a Eulerian momentum equation to compute the bulk velocity on the spatial mesh rather than using a Lagrangian particle discretization. As shown in Fig.~\ref{m1eigenfunction}, there is good agreement between these algorithms. Small differences close to the singular point ($r<0.1$) are due to statistical noise in the hybrid-PIC simulation, which is larger in this region due to the small cell volumes close to the singular point. The numerical profiles are also in agreement with the resistive-MHD theoretical eigenfunction.

The amplitudes of the density $||\delta n||(t) = ||n(t) - n(t=0)||$ and vector potential perturbations $||\delta A_z||(t) = ||A_z(t) - A_z(t=0)||$ are shown in the right panel of Fig.~\ref{m1eigenfunction} from the Hall-MHD (green) and hybrid-PIC (blue) simulations. The growth rate is in good agreement between the two codes, and these are both in agreement with the resistive-MHD linear theory (red). We note that the hybrid-PIC simulation is only run up to $t\approx 9.5\tau_0$. At this time in the non-linear regime, there are significant bulk flows across the singular point region causing the number of particles within these cells to decrease. The simulation then fails to converge due to the increased noise fluctuation level. We note that achieving the current simulation results was non-trivial, requiring the use of suitably tailored radial mesh packing profiles, macro-particle placement, and adjustment of particle weights to match the initial moment profiles. Future directions will explore the use of particle remapping and/or $\delta F$ methods in order to drastically reduce noise and allow the simulation to proceed to saturation. These are significant modifications to the algorithm and are outside of the scope of the present paper.

\section{Summary}

In this paper, we have derived a novel conservative scheme for the electromagnetic hybrid kinetic-ion fluid-electron plasma model by using implicit particle-in-cell methods. This work extends the work of Ref.~\cite{stanier19cart} -- which was derived for uniform Cartesian meshes -- to general curvilinear meshes by using a smooth and static map from logical to physical space. The equations for the electromagnetic fields and electron fluid quantities are discretized on a cell-centered mesh in logical space. For the kinetic-ion particles, the velocities are calculated in Cartesian space to avoid ficticious force terms in the equation of motion, while the particle positions are computed in logical space. The gather and scatter particle-mesh interpolations are performed using the Cartesian vector components at the logical space positions. These choices, along with the multi-rate implicit mid-point timestepping scheme used in Ref.~\cite{stanier19cart}, give the unique conservation properties of this algorithm. Sec.~\ref{energyconsv} demonstrates that the scheme discretely conserves the total (ion kinetic $+$ magnetic $+$ electron thermal) energy for general curvilinear meshes, provided suitable boundary conditions are used. Furthermore, these choices also give linear momentum conservation in the electrostatic limit as proved in Section~\ref{electrostaticmomentum}, and demonstrated numerically in Section~\ref{ESiaw}. For the full electromagnetic model, total momentum is conserved for tensor-packed meshes using a Cartesian coordinate basis (Sec.~\ref{planargrids} and verified numerically in Secs.~\ref{whistler}, \ref{islandcoal}). Although this property only applies to this restricted subset, these types of meshes are often used to efficiently study multiscale problems with thin boundary layers such as the magnetic island coalescence problem of Sec.~\ref{islandcoal}. In Sec.~\ref{numerical}, we also present a formal grid convergence study of the algorithm for simulating right-hand polarized whistler waves, which demonstrates second-order convergence for different types of meshes, and also present results from a challenging simulation of the $m=n=1$ mode of a plasma column on a 2D helically symmetric mesh geometry with non-periodic boundary conditions.

Future work on the kinetic-ion fluid-electron model will include the development of optimal and robust physics-based preconditioning strategies, and document the performance of the scheme. We will also explore noise reduction via $\delta F$ control-variate techniques.

\section*{Acknowledgements}
This work is supported by the Applied Scientific Computing Research (ASCR) program of the U.S. Department of Energy, and used resources provided by the Los Alamos National Laboratory Institutional Computing Program, which is supported by the U.S. Department of Energy National Nuclear Security Administration under Contract No. DE-AC52-06NA25396. 

\appendix
\section{Discrete vector identities in curvilinear coordinates}\label{sec:curvilinearidentities}

The vector product operations used in Sec.~\ref{sec:discrete} are defined in curvilinear form using index notation as

\begin{equation}\boldsymbol{a} \cdot \boldsymbol{b} = a^\alpha b_\alpha = a_\alpha b^\alpha,\end{equation}
\begin{equation}\left(\boldsymbol{a}\times \boldsymbol{b}\right)^\alpha = \frac{1}{J} \epsilon^{\alpha \beta \gamma} a_\beta b_\gamma,\end{equation}
\begin{equation}\left(\boldsymbol{a}\times \boldsymbol{b}\right)_\alpha = J \epsilon_{\alpha\beta\gamma}a^\beta b^\gamma,\end{equation}
where $\epsilon^{\alpha \beta \gamma}$, $\epsilon_{\alpha \beta \gamma}$ are Levi-Cevita symbols. The differential operators are defined as
\begin{equation}\boldsymbol{\nabla} \chi = \partial_\alpha \chi,\end{equation}
\begin{equation}\boldsymbol{\nabla} \cdot \boldsymbol{a} = \frac{1}{J} \partial_\alpha \left( J a^\alpha\right),\end{equation}
\begin{equation}\left(\boldsymbol{\nabla}\times \boldsymbol{a}\right)^\alpha = \frac{1}{J}\epsilon^{\alpha \beta \gamma}\partial_\beta a_\gamma.\end{equation}

\section{Boundary conditions}\label{sec:boundaryconditions}

The helical kink mode example problem discussed in Sec.~\ref{m1helical} uses non-periodic boundaries in the radial direction. These are a singular point boundary condition at $r=0$, and a conducting shell boundary condition at $r=1$. To simplify the treatment of these boundary conditions, we use Nearest Grid Point (NGP) interpolation for the radial direction in the scatter and gather operations of Eqs.~(\ref{scatterB},~\ref{scatterE},~\ref{densitymoment},~\ref{currentmoment}), while using second order in the periodic direction of the helical angle. This choice was made to achieve some noise reduction while ensuring that the particle-mesh interpolations only access cells interior to the domain boundaries.

\subsection{Conducting boundary condition}

For the particles we use a reflecting boundary condition
\begin{equation}\boldsymbol{v}_p \rightarrow \boldsymbol{v}_p -2\boldsymbol{v}_p\cdot \boldsymbol{\hat{n}}\boldsymbol{\hat{n}},\end{equation}
where $\boldsymbol{\hat{n}}$ is the normal vector to the boundary evaluated at the particle position. This conserves the particle energy, but not the momentum.

For the conducting wall we set $\boldsymbol{\hat{n}}\times \boldsymbol{E} = -\partial_t\boldsymbol{\hat{n}}\times \boldsymbol{A}  = 0$ and $\partial_t p_e = 0$. In cylindrical and helical coordinates this gives
\begin{equation}p_e(t) = p_e(t=0), \quad A_2(t) = A_2(t=0),\quad A_3(t) = A_3(t=0).\end{equation}


\subsection{Singular point boundary condition}

For mesh quantities, we note that our cell centered description means that no quantities are defined exactly on the singular point. For a radial mesh packing function that is symmetric about the singular point, the ghost cell (index $i=0$) is located at $r_0=-r_1$ where $r_1$ is the cell center of the first interior cell (index $i=1$). To set the singular point boundary conditions requires values to be set at this ghost cell. We follow Ref.~\cite{delzanno2008} and make use of the identity transformation for cylindrical/helical coordinates about the singular point $(r,\theta) \rightarrow (-r, \theta + \pi)$. Scalar quantities are thus filled as
\begin{equation}\chi(r_0,\theta) = \chi(r_1, \theta+\pi).\end{equation}

For our definition of basis functions in Sec.~\ref{sec:discrete}, the contravariant components of a vector $\boldsymbol{\Pi}$ can be expressed in cylindrical co-ordinates as $\Pi^1 = \Pi_r$, $\Pi^2 = \Pi_\theta/r$ and $\Pi^3 = \Pi_z$. The ghost cell values are found as follows $\Pi^1(r_0,\theta) = -\Pi^1(r_1, \theta+\pi)$, $\Pi^2(r_0,\theta) = \Pi^2(r_1, \theta+\pi)$, $\Pi^3(r_0,\theta) = \Pi^3(r_1, \theta+\pi)$. The covariant components are given by $\Pi_1=\Pi_r$, $\Pi_2=r\Pi_\theta$, $\Pi_3 = \Pi_z$, with the ghost cells being set in the same way. We note that this method is the same as in Ref.~\cite{delzanno2008} after accounting for differences in the basis vector definitions used in Sec.~\ref{sec:discrete}. 

\section{Conservative binomial smoothing}\label{conservativesmoothing}

Binomial smoothing is a useful technique for full-kinetic-PIC or hybrid-PIC schemes to reduce the effects of statistical noise due to a finite number of macro-particles~\cite{birdsall91}, while being computationally cheaper to apply than using higher-order shape functions (at least for the typical case where the number of particles-per-cell is large). The 3D smoothing operator is defined as $\textrm{SM}_g[Q_{g}] \equiv \textrm{SM}_k[\textrm{SM}_j[\textrm{SM}_i[Q_{ijk}]]]$, which is a composition of 1D smoothing operators defined as 
\begin{equation}\label{1dsmooth}\textrm{SM}_i\left[Q_{ijk}\right] = \frac{Q_{i-1jk} + 2Q_{ijk} + Q_{i+1jk}}{4}.\end{equation}

It is possible to do this smoothing in a conservative manner for the hybrid-PIC method, provided that it is done symmetrically to the electromagnetic field and moment quantities as defined in e.g., Ref.~\cite{stanier19cart}. This follows directly from the property of Eq.~(\ref{1dsmooth}), that $\sum_i A_{ijk} \textrm{SM}_i \left[B_{ijk}\right] = \sum_i  \textrm{SM}_i \left[A_{ijk}\right]B_{ijk}$ for suitable (e.g., periodic) boundary conditions. In a similar manner, it can be shown that the discrete conservation properties of Section~\ref{sec:conserve} continue to hold for curvilinear meshes when the smoothing is done as described below.

For the electromagnetic fields, the Cartesian vector representations of Eqs.~(\ref{scatterE}, \ref{scatterB}) are smoothed on the spatial mesh before scattering to the particles as
\begin{equation}\label{scatterEsmooth}\boldsymbol{E}_p^{*,\nu+1/2} = \sum_{g} \textrm{SM}_g\left[\boldsymbol{E}^{*,n+1/2}_{g}\right] S(\xi^\alpha_{g} - (\xi^{\alpha})_p^{\nu+1/2}),\end{equation}
\begin{equation}\label{scatterBsmooth}\boldsymbol{B}_p^{\nu+1/2} = \sum_{g} \textrm{SM}_g\left[\boldsymbol{B}^{n+1/2}_{g}\right] S(\xi^\alpha_{g} - (\xi^{\alpha})_p^{\nu+1/2}).\end{equation}

In addition, the Jacobian-weighted density and Cartesian momentum vectors of Eqs.~(\ref{densitymoment}, \ref{currentmoment}) are smoothed as
\begin{equation}\label{densitymomentsmooth} J_g n_{g}^{n+1/2} = \textrm{SM}_g\left[\frac{1}{\Delta V}\sum_p \frac{q_p}{e} w_p \frac{1}{\Delta t} \sum_{\nu=0}^{N_{\nu p} -1}S(\xi^\alpha_{g} - (\xi^{\alpha})_p^{\nu+1/2}) \Delta \tau_p^\nu\right]_g,\end{equation}
\begin{equation}\label{currentmomentsmooth} J_g (n\boldsymbol{u})_{g}^{n+1/2} = \textrm{SM}_g\left[\frac{1}{\Delta V}\sum_p \frac{q_p}{e} w_p \frac{1}{\Delta t} \sum_{\nu=0}^{N_{\nu p} -1} S(\xi^\alpha_{g} - (\xi^{\alpha})_p^{\nu+1/2}) \boldsymbol{v}_p^{\nu+1/2} \Delta \tau_p^\nu\right]_g.\end{equation}

We note that, as discussed in Ref.~\cite{stanier20cancel}, care must be taken to resolve the ion skin depth ($d_i$) when using higher-order particle shape functions or mesh-based smoothing, to avoid numerical dispersion errors due to the hybrid-PIC cancellation problem. Also, the compensation filter can be additionally applied in the same manner as Eqs.~(\ref{scatterEsmooth}-\ref{currentmomentsmooth}) to reduce the amount of attenuation caused by smoothing at long wavelengths ($k\Delta x\ll \pi$) where it is undesirable, see e.g., Refs~\cite{birdsall91,stanier19cart}.

\bibliographystyle{elsarticle-num}

\begin{thebibliography}{10}
\expandafter\ifx\csname url\endcsname\relax
  \def\url#1{\texttt{#1}}\fi
\expandafter\ifx\csname urlprefix\endcsname\relax\def\urlprefix{URL }\fi
\expandafter\ifx\csname href\endcsname\relax
  \def\href#1#2{#2} \def\path#1{#1}\fi

\bibitem{stanier19cart}
A.~Stanier, L.~Chacon, G.~Chen, A fully implicit, conservative, non-linear,
  electromagnetic hybrid particle-ion/fluid-electron algorithm, Journal of
  Computational Physics 376 (2019) 597--616.

\bibitem{priest2014}
E.~Priest, Magnetohydrodynamics of the Sun, Cambridge University Press, 2014.

\bibitem{schindler2006}
K.~Schindler, Physics of space plasma activity, Cambridge University Press,
  2006.

\bibitem{lipatov02}
A.~S. {Lipatov}, {The hybrid multiscale simulation technology: an introduction
  with application to astrophysical and laboratory plasmas}, Scientific
  Computation, Springer-Verlag Berlin Heidelberg, 2002.

\bibitem{byers78}
J.~Byers, B.~Cohen, W.~Condit, J.~Hanson,
  \href{http://www.sciencedirect.com/science/article/pii/0021999178900165}{Hybrid
  simulations of quasineutral phenomena in magnetized plasma}, Journal of
  Computational Physics 27~(3) (1978) 363 -- 396.
\newblock \href
  {http://dx.doi.org/https://doi.org/10.1016/0021-9991(78)90016-5}
  {\path{doi:https://doi.org/10.1016/0021-9991(78)90016-5}}.
\newline\urlprefix\url{http://www.sciencedirect.com/science/article/pii/0021999178900165}

\bibitem{hewett78}
D.~W. {Hewett}, C.~W. {Nielson}, {A Multidimensional Quasineutral Plasma
  Simulation Model}, Journal of Computational Physics 29~(2) (1978) 219--236.
\newblock \href {http://dx.doi.org/10.1016/0021-9991(78)90153-5}
  {\path{doi:10.1016/0021-9991(78)90153-5}}.

\bibitem{winske03}
D.~{Winske}, L.~{Yin}, N.~{Omidi}, H.~{Karimabadi}, K.~{Quest}, {Hybrid
  Simulation Codes: Past, Present and Future - A Tutorial}, in:
  J.~{B{\"u}chner}, C.~{Dum}, M.~{Scholer} (Eds.), Space Plasma Simulation,
  Vol. 615 of Lecture Notes in Physics, Berlin Springer Verlag, 2003, pp.
  136--165.

\bibitem{belova1997}
E.~Belova, R.~Denton, A.~Chan,
  \href{https://www.sciencedirect.com/science/article/pii/S0021999197957387}{Hybrid
  simulations of the effects of energetic particles on low-frequency mhd
  waves}, Journal of Computational Physics 136~(2) (1997) 324--336.
\newblock \href {http://dx.doi.org/https://doi.org/10.1006/jcph.1997.5738}
  {\path{doi:https://doi.org/10.1006/jcph.1997.5738}}.
\newline\urlprefix\url{https://www.sciencedirect.com/science/article/pii/S0021999197957387}

\bibitem{drake2019}
J.~Drake, H.~Arnold, M.~Swisdak, J.~Dahlin, A computational model for exploring
  particle acceleration during reconnection in macroscale systems, Physics of
  Plasmas 26~(1) (2019) 012901.

\bibitem{holderied2021}
F.~Holderied, S.~Possanner, X.~Wang,
  \href{https://www.sciencedirect.com/science/article/pii/S0021999121000358}{Mhd-kinetic
  hybrid code based on structure-preserving finite elements with
  particles-in-cell}, Journal of Computational Physics 433 (2021) 110143.
\newblock \href {http://dx.doi.org/https://doi.org/10.1016/j.jcp.2021.110143}
  {\path{doi:https://doi.org/10.1016/j.jcp.2021.110143}}.
\newline\urlprefix\url{https://www.sciencedirect.com/science/article/pii/S0021999121000358}

\bibitem{amano2018}
T.~Amano, A generalized quasi-neutral fluid-particle hybrid plasma model and
  its application to energetic-particle-magnetohydrodynamics hybrid simulation,
  Journal of Computational Physics 366 (2018) 366--385.

\bibitem{birdsall91}
C.~K. {Birdsall}, A.~B. {Langdon}, {Plasma Physics via Computer Simulation},
  McGraw-Hill, New York, 1991.

\bibitem{karimabadi04}
H.~{Karimabadi}, D.~{Krauss-Varban}, J.~D. {Huba}, H.~X. {Vu}, {On magnetic
  reconnection regimes and associated three-dimensional asymmetries: Hybrid,
  Hall-less hybrid, and Hall-MHD simulations}, Journal of Geophysical Research
  (Space Physics) 109 (2004) A09205.
\newblock \href {http://dx.doi.org/10.1029/2004JA010478}
  {\path{doi:10.1029/2004JA010478}}.

\bibitem{stanier15prl}
A.~{Stanier}, W.~{Daughton}, L.~{Chac{\'o}n}, H.~{Karimabadi}, J.~{Ng}, Y.-M.
  {Huang}, A.~{Hakim}, A.~{Bhattacharjee}, {Role of Ion Kinetic Physics in the
  Interaction of Magnetic Flux Ropes}, Physical Review Letters 115~(17) (2015)
  175004.
\newblock \href {http://dx.doi.org/10.1103/PhysRevLett.115.175004}
  {\path{doi:10.1103/PhysRevLett.115.175004}}.

\bibitem{le2019}
A.~Le, A.~Stanier, W.~Daughton, J.~Ng, J.~Egedal, W.~D. Nystrom, R.~Bird,
  Three-dimensional stability of current sheets supported by electron pressure
  anisotropy, Physics of Plasmas 26~(10) (2019) 102114.

\bibitem{winske1985}
D.~Winske, Hybrid simulation codes with application to shocks and upstream
  waves, Space Science Reviews 42~(1-2) (1985) 53--66.

\bibitem{weidl2016}
M.~S. Weidl, D.~Winske, F.~Jenko, C.~Niemann, Hybrid simulations of a parallel
  collisionless shock in the large plasma device, Physics of Plasmas 23~(12)
  (2016) 122102.

\bibitem{le2021}
A.~Le, D.~Winske, A.~Stanier, W.~Daughton, M.~Cowee, B.~Wetherton, F.~Guo,
  Astrophysical explosions revisited: collisionless coupling of debris to
  magnetized plasma, Journal of Geophysical Research: Space Physics (2021)
  e2021JA029125.

\bibitem{franci2015}
L.~Franci, S.~Landi, L.~Matteini, A.~Verdini, P.~Hellinger, High-resolution
  hybrid simulations of kinetic plasma turbulence at proton scales, The
  Astrophysical Journal 812~(1) (2015) 21.

\bibitem{cerri2016}
S.~S. Cerri, F.~Califano, F.~Jenko, D.~Told, F.~Rincon, Subproton-scale
  cascades in solar wind turbulence: driven hybrid-kinetic simulations, The
  Astrophysical Journal Letters 822~(1) (2016) L12.

\bibitem{le2018wavelet}
A.~Le, V.~Roytershteyn, H.~Karimabadi, A.~Stanier, L.~Chacon, K.~Schneider,
  Wavelet methods for studying the onset of strong plasma turbulence, Physics
  of Plasmas 25~(12) (2018) 122310.

\bibitem{muller2012}
J.~M{\"u}ller, S.~Simon, Y.-C. Wang, U.~Motschmann, D.~Heyner, J.~Sch{\"u}le,
  W.-H. Ip, G.~Kleindienst, G.~J. Pringle, Origin of mercury’s double
  magnetopause: 3d hybrid simulation study with aikef, Icarus 218~(1) (2012)
  666--687.

\bibitem{karimabadi14}
H.~{Karimabadi}, V.~{Roytershteyn}, H.~X. {Vu}, Y.~A. {Omelchenko},
  J.~{Scudder}, W.~{Daughton}, A.~{Dimmock}, K.~{Nykyri}, M.~{Wan},
  D.~{Sibeck}, M.~{Tatineni}, A.~{Majumdar}, B.~{Loring}, B.~{Geveci}, {The
  link between shocks, turbulence, and magnetic reconnection in collisionless
  plasmas}, Physics of Plasmas 21~(6) (2014) 062308.
\newblock \href {http://dx.doi.org/10.1063/1.4882875}
  {\path{doi:10.1063/1.4882875}}.

\bibitem{omidi2017}
N.~Omidi, G.~Collinson, D.~Sibeck, Structure and properties of the foreshock at
  venus, Journal of Geophysical Research: Space Physics 122~(10) (2017)
  10--275.

\bibitem{ng2021bursty}
J.~Ng, L.-J. Chen, Y.~Omelchenko, Bursty magnetic reconnection at the earth's
  magnetopause triggered by high-speed jets, Physics of Plasmas 28~(9) (2021)
  092902.

\bibitem{harned82}
D.~S. {Harned}, {Quasineutral hybrid simulation of macroscopic plasma
  phenomena}, Journal of Computational Physics 47 (1982) 452--462.
\newblock \href {http://dx.doi.org/10.1016/0021-9991(82)90094-8}
  {\path{doi:10.1016/0021-9991(82)90094-8}}.

\bibitem{winske86}
D.~{Winske}, K.~B. {Quest}, {Electromagnetic ion beam instabilities -
  Comparison of one- and two-dimensional simulations}, Journal of Geophysical
  Research 91 (1986) 8789--8797.
\newblock \href {http://dx.doi.org/10.1029/JA091iA08p08789}
  {\path{doi:10.1029/JA091iA08p08789}}.

\bibitem{kunz13}
M.~W. {Kunz}, G.~{Lesur}, {Magnetic self-organization in Hall-dominated
  magnetorotational turbulence}, Monthly Notices of the Royal Astronomical
  Society 434 (2013) 2295--2312.
\newblock \href {http://arxiv.org/abs/1306.5887} {\path{arXiv:1306.5887}},
  \href {http://dx.doi.org/10.1093/mnras/stt1171}
  {\path{doi:10.1093/mnras/stt1171}}.

\bibitem{winske88}
D.~{Winske}, K.~B. {Quest}, {Magnetic field and density fluctuations at
  perpendicular supercritical collisionless shocks}, Journal of Geophysical
  Research 93 (1988) 9681--9693.
\newblock \href {http://dx.doi.org/10.1029/JA093iA09p09681}
  {\path{doi:10.1029/JA093iA09p09681}}.

\bibitem{matthews94}
A.~P. Matthews,
  \href{//www.sciencedirect.com/science/article/pii/S0021999184710849}{Current
  advance method and cyclic leapfrog for 2d multispecies hybrid plasma
  simulations}, Journal of Computational Physics 112~(1) (1994) 102 -- 116.
\newblock \href {http://dx.doi.org/http://dx.doi.org/10.1006/jcph.1994.1084}
  {\path{doi:http://dx.doi.org/10.1006/jcph.1994.1084}}.
\newline\urlprefix\url{//www.sciencedirect.com/science/article/pii/S0021999184710849}

\bibitem{fujimoto90}
M.~Fujimoto, Instabilities in the magnetopause velocity shear layer, Ph.D.
  thesis, Univ. of Tokyo, Tokyo (1990).

\bibitem{thomas90}
V.~A. Thomas, D.~Winske, N.~Omidi,
  \href{http://dx.doi.org/10.1029/JA095iA11p18809}{Re-forming supercritical
  quasi-parallel shocks: 1. one- and two-dimensional simulations}, Journal of
  Geophysical Research: Space Physics 95~(A11) (1990) 18809--18819.
\newblock \href {http://dx.doi.org/10.1029/JA095iA11p18809}
  {\path{doi:10.1029/JA095iA11p18809}}.
\newline\urlprefix\url{http://dx.doi.org/10.1029/JA095iA11p18809}

\bibitem{swift95}
D.~W. Swift, \href{http://dx.doi.org/10.1029/94GL03082}{Use of a hybrid code to
  model the earth's magnetosphere}, Geophysical Research Letters 22~(3) (1995)
  311--314.
\newblock \href {http://dx.doi.org/10.1029/94GL03082}
  {\path{doi:10.1029/94GL03082}}.
\newline\urlprefix\url{http://dx.doi.org/10.1029/94GL03082}

\bibitem{hewett80}
D.~W. {Hewett}, {A global method of solving the electron-field equations in a
  zero-inertia-electron-hybrid plasma simulation code}, Journal of
  Computational Physics 38 (1980) 378--395.
\newblock \href {http://dx.doi.org/10.1016/0021-9991(80)90155-2}
  {\path{doi:10.1016/0021-9991(80)90155-2}}.

\bibitem{amano14}
T.~{Amano}, K.~{Higashimori}, K.~{Shirakawa}, {A robust method for handling low
  density regions in hybrid simulations for collisionless plasmas}, Journal of
  Computational Physics 275 (2014) 197--212.
\newblock \href {http://arxiv.org/abs/1406.6613} {\path{arXiv:1406.6613}},
  \href {http://dx.doi.org/10.1016/j.jcp.2014.06.048}
  {\path{doi:10.1016/j.jcp.2014.06.048}}.

\bibitem{sturdevant17}
B.~J. {Sturdevant}, Y.~{Chen}, S.~E. {Parker}, {Low frequency fully kinetic
  simulation of the toroidal ion temperature gradient instability}, Physics of
  Plasmas 24~(8) (2017) 081207.
\newblock \href {http://dx.doi.org/10.1063/1.4999945}
  {\path{doi:10.1063/1.4999945}}.

\bibitem{rambo95}
P.~W. {Rambo}, {Finite-Grid Instability in Quasineutral Hybrid Simulations},
  Journal of Computational Physics 118 (1995) 152--158.
\newblock \href {http://dx.doi.org/10.1006/jcph.1995.1086}
  {\path{doi:10.1006/jcph.1995.1086}}.

\bibitem{higginson19}
D.~Higginson, P.~Amendt, N.~Meezan, W.~Riedel, H.~Rinderknecht, S.~Wilks,
  G.~Zimmerman, Hybrid particle-in-cell simulations of laser-driven plasma
  interpenetration, heating, and entrainment, Physics of Plasmas 26~(11) (2019)
  112107.

\bibitem{leblond11phd}
D.~Leblond, Simulation des plasmas de tokamak avec xtor: r{\'e}gimes des dents
  de scie et {\'e}volution vers une mod{\'e}lisation cin{\'e}tique des ions,
  Ph.D. thesis (2011).

\bibitem{dyadechkin13}
S.~Dyadechkin, E.~Kallio, R.~Jarvinen, A new 3-d spherical hybrid model for
  solar wind interaction studies, Journal of Geophysical Research: Space
  Physics 118~(8) (2013) 5157--5168.

\bibitem{muller2011aikef}
J.~M{\"u}ller, S.~Simon, U.~Motschmann, J.~Sch{\"u}le, K.-H. Glassmeier, G.~J.
  Pringle, Aikef: Adaptive hybrid model for space plasma simulations, Computer
  Physics Communications 182~(4) (2011) 946--966.

\bibitem{colella10}
P.~Colella, P.~C. Norgaard, Controlling self-force errors at refinement
  boundaries for amr-pic, Journal of Computational Physics 229~(4) (2010)
  947--957.

\bibitem{stanier20cancel}
A.~Stanier, L.~Chacon, A.~Le, A cancellation problem in hybrid particle-in-cell
  schemes due to finite particle size, Journal of Computational Physics 420
  (2020) 109705.

\bibitem{chacon04}
L.~{Chac{\'o}n}, {A non-staggered, conservative, div(B)=0, finite-volume scheme
  for 3D implicit extended magnetohydrodynamics in curvilinear geometries},
  Computer Physics Communications 163 (2004) 143--171.
\newblock \href {http://dx.doi.org/10.1016/j.cpc.2004.08.005}
  {\path{doi:10.1016/j.cpc.2004.08.005}}.

\bibitem{chen15}
G.~{Chen}, L.~{Chac{\'o}n}, {A multi-dimensional, energy- and
  charge-conserving, nonlinearly implicit, electromagnetic Vlasov-Darwin
  particle-in-cell algorithm}, Computer Physics Communications 197 (2015)
  73--87.
\newblock \href {http://arxiv.org/abs/1503.01169} {\path{arXiv:1503.01169}},
  \href {http://dx.doi.org/10.1016/j.cpc.2015.08.008}
  {\path{doi:10.1016/j.cpc.2015.08.008}}.

\bibitem{cohen82}
B.~I. Cohen, R.~P. Freis, V.~Thomas,
  \href{http://www.sciencedirect.com/science/article/pii/0021999182901085}{Orbit-averaged
  implicit particle codes}, Journal of Computational Physics 45~(3) (1982) 345
  -- 366.
\newblock \href
  {http://dx.doi.org/https://doi.org/10.1016/0021-9991(82)90108-5}
  {\path{doi:https://doi.org/10.1016/0021-9991(82)90108-5}}.
\newline\urlprefix\url{http://www.sciencedirect.com/science/article/pii/0021999182901085}

\bibitem{liseikin2017}
V.~D. Liseikin, Grid generation methods, Springer, 2017.

\bibitem{hirt68}
C.~W. {Hirt}, {Heuristic Stability Theory for Finite-Difference Equations},
  Journal of Computational Physics 2 (1968) 339--355.
\newblock \href {http://dx.doi.org/10.1016/0021-9991(68)90041-7}
  {\path{doi:10.1016/0021-9991(68)90041-7}}.

\bibitem{knoll04}
D.~A. {Knoll}, D.~E. {Keyes}, {Jacobian-free Newton-Krylov methods: a survey of
  approaches and applications}, Journal of Computational Physics 193 (2004)
  357--397.
\newblock \href {http://dx.doi.org/10.1016/j.jcp.2003.08.010}
  {\path{doi:10.1016/j.jcp.2003.08.010}}.

\bibitem{saad93}
Y.~Saad, \href{https://doi.org/10.1137/0914028}{A flexible inner-outer
  preconditioned gmres algorithm}, SIAM Journal on Scientific Computing 14~(2)
  (1993) 461--469.
\newblock \href {http://arxiv.org/abs/https://doi.org/10.1137/0914028}
  {\path{arXiv:https://doi.org/10.1137/0914028}}, \href
  {http://dx.doi.org/10.1137/0914028} {\path{doi:10.1137/0914028}}.
\newline\urlprefix\url{https://doi.org/10.1137/0914028}

\bibitem{sydora1999}
R.~Sydora, Low-noise electromagnetic and relativistic particle-in-cell plasma
  simulation models, Journal of computational and applied mathematics 109~(1-2)
  (1999) 243--259.

\bibitem{hammersley2013}
J.~Hammersley, Monte carlo methods, Springer Science \& Business Media, 2013.

\bibitem{finn1977}
J.~M. Finn, P.~Kaw, Coalescence instability of magnetic islands, The Physics of
  Fluids 20~(1) (1977) 72--78.

\bibitem{pritchett1992coalescence}
P.~Pritchett, The coalescence instability in collisionless plasmas, Physics of
  Fluids B: Plasma Physics 4~(10) (1992) 3371--3381.

\bibitem{karimabadi11}
H.~{Karimabadi}, J.~{Dorelli}, V.~{Roytershteyn}, W.~{Daughton},
  L.~{Chac{\'o}n}, Phys. Rev. Lett. 107~(2) (2011) 025002.
\newblock \href {http://dx.doi.org/10.1103/PhysRevLett.107.025002}
  {\path{doi:10.1103/PhysRevLett.107.025002}}.

\bibitem{ng15}
J.~{Ng}, Y.-M. {Huang}, A.~{Hakim}, A.~{Bhattacharjee}, A.~{Stanier},
  W.~{Daughton}, L.~{Wang}, K.~{Germaschewski}, {The island coalescence
  problem: Scaling of reconnection in extended fluid models including
  higher-order moments}, Physics of Plasmas 22~(11) (2015) 112104.
\newblock \href {http://arxiv.org/abs/1511.00741} {\path{arXiv:1511.00741}},
  \href {http://dx.doi.org/10.1063/1.4935302} {\path{doi:10.1063/1.4935302}}.

\bibitem{stanier17}
A.~Stanier, W.~Daughton, A.~N. Simakov, L.~Chacon, A.~Le, H.~Karimabadi, J.~Ng,
  A.~Bhattacharjee, The role of guide field in magnetic reconnection driven by
  island coalescence, Physics of Plasmas 24~(2) (2017) 022124.

\bibitem{makwana2018}
K.~Makwana, R.~Keppens, G.~Lapenta, Two-way coupled mhd-pic simulations of
  magnetic reconnection in magnetic island coalescence, in: Journal of Physics:
  Conference Series, Vol. 1031, IOP Publishing, 2018, p. 012019.

\bibitem{allmann2018}
F.~Allmann-Rahn, T.~Trost, R.~Grauer, Temperature gradient driven heat flux
  closure in fluid simulations of collisionless reconnection, Journal of Plasma
  Physics 84~(3).

\bibitem{fadeev1965}
V.~Fadeev, I.~Kvabtskhava, N.~Komarov, Self-focusing of local plasma currents,
  Nuclear fusion 5~(3) (1965) 202.

\bibitem{stanier15b}
A.~{Stanier}, A.~N. {Simakov}, L.~{Chac{\'o}n}, W.~{Daughton}, {Fluid vs.
  kinetic magnetic reconnection with strong guide fields}, Physics of Plasmas
  22~(10) (2015) 101203.
\newblock \href {http://dx.doi.org/10.1063/1.4932330}
  {\path{doi:10.1063/1.4932330}}.

\bibitem{bennett1934}
W.~H. Bennett,
  \href{https://link.aps.org/doi/10.1103/PhysRev.45.890}{Magnetically
  self-focussing streams}, Phys. Rev. 45 (1934) 890--897.
\newblock \href {http://dx.doi.org/10.1103/PhysRev.45.890}
  {\path{doi:10.1103/PhysRev.45.890}}.
\newline\urlprefix\url{https://link.aps.org/doi/10.1103/PhysRev.45.890}

\bibitem{chen2021}
G.~Chen, L.~Chac{\'o}n, T.~B. Nguyen, An unsupervised machine-learning
  checkpoint-restart algorithm using gaussian mixtures for particle-in-cell
  simulations, Journal of Computational Physics 436 (2021) 110185.

\bibitem{mishchenko12}
A.~Mishchenko, A.~Zocco, Global gyrokinetic particle-in-cell simulations of
  internal kink instabilities, Physics of Plasmas 19~(12) (2012) 122104.

\bibitem{delzanno2008}
G.~L. Delzanno, L.~Chac{\'o}n, J.~M. Finn, Electrostatic mode associated with
  the pinch velocity in reversed field pinch simulations, Physics of Plasmas
  15~(12) (2008) 122102.

\end{thebibliography}







\end{document}